\documentclass[11pt]{article}
\usepackage{jheppub}
\usepackage{amsmath, amssymb, amsfonts, pifont}
\usepackage{etoolbox}
\makeatletter
\patchcmd{\maketitle}{\@fpheader}{}{}{}
\makeatother

\newcommand{\be}{\begin{equation}}
\newcommand{\ee}{\end{equation}}
\newcommand{\beq}{\begin{equation}}
\newcommand{\eeq}{\end{equation}}
\newcommand{\bea}{\begin{eqnarray}}
\newcommand{\eea}{\end{eqnarray}}
\newcommand{\ba}{\begin{array}}
\newcommand{\ea}{\end{array}}
\newcommand{\nn}{\nonumber}
\newcommand{\tr}{\mathrm{tr}\,}
\newcommand{\IC}{\mathbb{C}}
\newcommand{\IP}{\mathbb{P}}
\newcommand{\IH}{\mathbb{H}}
\newcommand{\IO}{\mathbb{O}}

\newcommand\cg{\mathfrak{g}}
\newcommand\kM{\mathfrak{M}}
\newcommand\kN{\mathfrak{N}}
\newcommand\CB{{\cal B}}
\newcommand{\cM}{{\cal M}}
\newcommand{\cN}{{\cal N}}

\newcommand{\eB}{{\tilde{e}}} %

\def\eps{\epsilon^{\alpha \beta}}
\def\barH{\overline{H}}
\newcommand{\gen}[1]{\langle #1 \rangle}

\newcommand{\IZ}{\mathbb{Z}}
\newcommand{\IR}{\mathbb{R}}

\newcommand{\comment}[1]{}

\renewcommand{\baselinestretch}{1.3}

\newcommand{\setall}{\setcounter{equation}{0}}

\newtheorem{theorem}	{Theorem}	[section]
\newtheorem{fact} [theorem] {Fact}
\newcommand{\Hom}{\mathop{\rm Hom}\nolimits}
\newcommand{\OO}{{\mathcal O}}
\newcommand{\coker}{\mathop{\rm coker}\nolimits}
\newcommand{\qed}{\hfill{$\blacksquare$}}

\begin{document}

\preprint{WITS-CTP-143}

\title{The Geometry of Generations}
\author[a,b,c]{Yang-Hui He,}
\author[d]{Vishnu Jejjala,}
\author[a]{Cyril Matti,}
\author[e,f]{Brent D.\ Nelson,}
\author[g]{Michael Stillman}
\affiliation[a]{Department of Mathematics, City University, London, EC1V 0HB, UK}
\affiliation[b]{School of Physics, NanKai University, Tianjin, 300071, P.R.\ China}
\affiliation[c]{Merton College, University of Oxford, OX1 4JD, UK}
\affiliation[d]{Centre for Theoretical Physics, NITheP, and School of Physics,
University of the Witwatersrand, Johannesburg, WITS 2050, South Africa}
\affiliation[e]{Department of Physics, Northeastern University, Boston, MA 02115, USA}
\affiliation[f]{ICTP, Strada Costiera 11, Trieste 34014, Italy}
\affiliation[g]{Department of Mathematics, Cornell University, Ithaca, NY 14853-4201, USA}
\emailAdd{hey@maths.ox.ac.uk}
\emailAdd{vishnu@neo.phys.wits.ac.za}
\emailAdd{cyril.matti.1@city.ac.uk}
\emailAdd{b.nelson@neu.edu}
\emailAdd{mike@math.cornell.edu}
\begin{abstract}{
We present an intriguing and precise interplay between algebraic geometry and the phenomenology of generations of particles. Using the electroweak sector of the MSSM as a testing ground, we compute the moduli space of vacua as an algebraic variety for multiple generations of Standard Model matter and Higgs doublets.  The space is shown to have Calabi--Yau, Grassmannian, and toric signatures which sensitively depend on the number of generations of leptons, as well as inclusion of Majorana mass terms for right-handed neutrinos. We speculate as to why three generations is special. 
}
\end{abstract}
\maketitle

\newpage

\section{Introduction}\label{intro}\setall

The Standard Model of particle physics is an incomplete theory of the gauge interactions.
We expect that the physics which extends the Standard Model at energies above $1$--$10$ TeV invokes supersymmetry and derives from some higher energy theory that also incorporates gravity.
A key property of any quantum field theory is its vacuum and in the context of supersymmetric gauge theories, the vacuum possesses interesting structure.
This is because the supersymmetric vacuum is the solution of F-flatness and D-flatness conditions.
Generically, this is a continuous manifold parametrized by the gauge invariant operators (GIOs) of the theory.
Importantly, this {\it vacuum moduli space} is an algebraic variety, which can have intricate geometric properties.
The topology and algebraic geometry of the vacuum is coextensive with phenomenology~\cite{Gray:2005sr,Gray:2006jb}.
Exploring the structure of the vacuum therefore provides a low energy window into deducing how certain theories of phenomenological interest can both encode and be guided by interesting geometry.

The most na\"{\i}ve extension of known particle physics is the MSSM, which expands the Higgs sector of the theory by introducing separate $SU(2)_L$ doublets for up-type quark and down-type quark Yukawa couplings.
The vacuum geometry of this theory, or related theories like the NMSSM, is not known, even though it has existed as a computational challenge to the community for many decades~\cite{Gherghetta:1995dv,lt,Buccella:1982nx,Gatto:1986bt,Procesi:hr}.
This is because the vacuum moduli spaces of ${\cal N}=1$ theories are expressed as relations between the generators of the GIOs, which are monomials in the superfields of the theory.
The minimal list contains $991$ generators for GIOs in the MSSM~\cite{Gherghetta:1995dv}.
These are not fully independent and are related by the $49$ F-term equations for the component matter superfields in the theory.
Although this effort motivated~\cite{Gray:2005sr,Gray:2006jb}, solving for the vacuum of the full MSSM was then beyond our reach.

While the exact vacuum geometry remains unknown, we can ask and hope to answer a different class of questions.
We know, for instance, that for generic numbers of flavors $N_f$ and colors $N_c$, the vacuum moduli space of supersymmetric QCD is a Calabi--Yau manifold~\cite{Gray:2008yu}.
Does this property extend to the vacuum geometry of the MSSM?
Is there something special, geometrically speaking, about the particle content that we see experimentally?
Why are there three generations of matter fields at low energies?
It is difficult to imagine questions that are more pressing, especially from the point of view of string theory, which purports to be a fundamental theory. In this case, the initial conditions that describe the Standard Model are, in fact, the result of some vacuum selection principle.

Theoretical physicists should be proceeding from low energy data only and be working from the ground up in order to establish a principle for understanding string vacua. This is in some sense orthogonal to the traditional techniques of gauge invariance and discrete symmetries.
A first attempt to define what such a program might look like is given in~\cite{Gray:2005sr,Gray:2006jb}.
Since that time, advances in computational algebraic geometry software, as well as overall advances in computation, have allowed us to probe further than could have been conceived eight years ago.
The first paper to re-address this fundamental problem appeared recently~\cite{He:2014loa}.
The current article seeks to build on these advances to explore the electroweak sector of the MSSM in the broadest possible context.
Fortified by the discovery of interesting geometry encoded in the vacuum moduli space of the MSSM, we wish to find out whether geometry can say anything new about the nature of generations of particles.
Intriguingly, it does.

The paper is organized as follows.
In Section~\ref{sec:two}, we review how to compute the vacuum moduli space of a supersymmetric theory.
This allows us to set notation and establish our conventions.
In Section~\ref{s:mutli} we present results obtained from considering a minimal renormalizable superpotential and various numbers of particle flavors.
We explicitly describe the vacuum geometry for the cases $N_f = 2, 3, 4, 5$.
In Section~\ref{sec:four}, we consider multiple generations of Higgs fields for this minimal superpotential.
In Section~\ref{sec:five}, we then move on to theories with right-handed neutrinos fields with Majorana mass terms and then without.
We give the vacuum geometry for the cases $N_f = 2, 3, 4$, as well as a general description for general $N_f$ in the case without Majorana mass terms.
We conclude in Section~\ref{conclusion}.
Appendices~\ref{GIOs}, \ref{toric} and \ref{ap:toricCY} contain complementary information about the full MSSM GIO content and about the method used to obtain toric diagrams from binomial ideals.

\section{MSSM Vacuum Moduli Space}\label{sec:two}\setall
We begin by reminding the reader of the algorithm with which we explicitly calculate the vacuum moduli space of supersymmetric gauge theories from the point of view of computational algebraic geometry, focusing on the MSSM.
First, we introduce the matter content and the superpotential and then we summarize the algorithm.
\subsection{F-terms and D-terms}

In order to set the scene and specify our notation, let us briefly review the context of four-dimensional supersymmetric gauge theories and the Minimal Supersymmetric Standard Model (MSSM) field content.

A general $\cN=1$ globally supersymmetric action in four dimensions is given by
\be
\label{action}
S = \int d^4x\ \left[ \int d^4\theta\ \Phi_i^\dagger e^V \Phi_i +
    \left( \frac{1}{4g^2} \int d^2\theta\ \tr{W_\alpha W^\alpha} +
    \int d^2\theta\ W(\Phi) + {\rm h.c.} \right) \right],
\ee
where $\Phi_i$ are chiral superfields, $V$ is a vector superfield, $W_\alpha$ are chiral spinor superfields, and the superpotential $W$ is a holomorphic function of the superfields $\Phi_i$.
Each of these objects transforms under the gauge group $G$ of the theory: $\Phi_i$ under some representation $R_i$ and $V$ in the Lie algebra $\cg$.
The chiral spinor superfields are the gauge field strength and are given by $W_\alpha = i\overline{D}^2 e^{-V} D_\alpha e^V$.

The vacuum of the theory consists of $\phi_{i 0}$, the vacuum expectation values of the scalar components of the superfields $\Phi_i$ that provide a simultaneous solution to the F-term equations
\be
\label{fterm}
\left. \frac{\partial W(\phi)}{\partial \phi_i}
\right|_{\phi_i=\phi_{i 0}} = 0 \;
\ee
and the D-term equations
\be
\label{dterm}
D^A = \sum_i \phi_{i 0}^\dagger\, T^A\, \phi_{i 0} = 0\; ,
\ee
where $T^A$ are generators of the gauge group in the adjoint representation, and we have chosen the Wess--Zumino gauge.

The MSSM fixes the gauge group $G = SU(3)_C\times SU(2)_L \times U(1)_Y$.
We will adopt the notation given in Table~\ref{conv} for the indices and the field content of the theory.
For the moment, we do not consider right-handed neutrinos, which are gauge singlets.

\begin{table}[h]
{\begin{center}
\begin{tabular}{ccc}
\begin{tabular}{|c|c|}\hline
INDICES & \\ \hline \hline
$i,j,k,l = 1,2,\ldots, N_{f}$ & Flavor (family) indices \\
$a,b,c, d = 1,2,3$ & $SU(3)_C$ color indices \\
$\alpha, \beta, \gamma, \delta = 1,2 $ & $SU(2)_L$ indices \\
\hline
\end{tabular}
&
\begin{tabular}{|c|c|}\hline
FIELDS & \\ \hline \hline
$Q_{a, \alpha}^i$ & $SU(2)_L$ doublet quarks \\
$u_a^i$ & $SU(2)_L$ singlet up-quarks \\
$d_a^i$ & $SU(2)_L$ singlet down-quarks \\
$L_{\alpha}^i$  & $SU(2)_L$ doublet leptons \\
$e^i$ & $SU(2)_L$ singlet leptons \\
$H_\alpha$ & up-type Higgs \\
$\barH_\alpha$ & down-type Higgs \\ \hline
\end{tabular}
\end{tabular}
\end{center}}
{\caption{\label{conv} {\sf Indices and field content conventions for the MSSM}.}}
\end{table}

The corresponding minimal renormalizable superpotential is
\bea\label{renorm}
W_{\rm minimal} = \; C^0 \sum_{\alpha, \beta} H_\alpha \barH_\beta \eps + \sum_{i,j} C^1_{ij} \sum_{\alpha, \beta, a} Q^i_{a,\alpha} u^j_a H_\beta \eps \nonumber \\
 +\sum_{i,j} C^2_{ij} \sum_{\alpha, \beta, a} Q^i_{a,\alpha} d^j_a \barH_\beta \eps + \sum_{i,j} C^3_{ij} e^i \sum_{\alpha, \beta} L^j_{\alpha} \barH_\beta \eps \; ,
\eea
where $C$ designates coupling constants and $\eps$ is the totally antisymmetric tensor.
These are the minimal terms consistent with assigning masses to the particles in the theory.
All of these terms respect R-parity.
The problem of finding the vacuum moduli space of the theory thus reduces to solving~\eqref{fterm} and~\eqref{dterm} for the above superpotential.

\subsection{Computational Algorithm}
\label{sec:algorithm}

Algebraic geometry has proven a useful and powerful tool to tackle problems in gauge fields theories, not least the challenge of providing a mathematical description of vacuum moduli spaces~\cite{book}.
The problem of solving~\eqref{fterm} and~\eqref{dterm} is equivalent to the elimination algorithm detailed below.

Let us denote the gauge invariant operators (GIOs) by $r_j( \{ \phi_i \})$.
The full list of generators for the MSSM GIOs is given in Appendix~\ref{GIOs}.\footnote{
This table was already presented in~\cite{Gray:2006jb} and stems from earlier work of~\cite{Gherghetta:1995dv}.
Here, we have corrected some minor typographical errors with respect to the indices.}
The description of the moduli space of ${\cal N}=1$ theories as the symplectic quotient of the space of F-flat field configurations by the complexified gauge group $G^C$ is well known~\cite{lt,Buccella:1982nx,Gatto:1986bt,Procesi:hr,witten93}.
Our goal is to provide an efficient methodology for implementing this result.

Let us consider the ideal
\begin{equation}
\left\langle
\frac{\partial {W}}{\partial \phi_i} \; ,
y_j - r_j( \{ \phi_i \})
\right\rangle
\subset
R = \IC[\phi_{i=1, \ldots, n}, y_{j = 1, \ldots, k}] \ ,
\end{equation}
where $y_i$ are additional variables.
Then, eliminating all variables $\phi_i$ of this ideal will give an ideal in terms of the variables $y_i$ in the polynomial ring $S=\IC[y_{j = 1, \ldots, k}]$ only.
The result will be the vacuum moduli space as an algebraic variety in $S$.
From the standpoint of algebraic geometry, the above prescription amounts to finding the image of a map from the quotient ring
\begin{equation} \mathcal{F} = \frac{\IC[\phi_{i=1, \ldots, n}]}{\langle
\frac{\partial {W}}{\partial \phi_i} \rangle} \label{quotring} \end{equation}
to the ring $S=\IC[y_{j = 1, \ldots, k}]$.\footnote{
It should be noted that, geometrically, the map goes from the algebraic variety to the $k$-affine space. However, the ring map goes in the other direction, from the ring $S$ to the quotient ring $\mathcal{F}$.
}
(See~\cite{Gray:2009fy,Hauenstein:2012xs} for further details.)

This algorithm can be summarized in the following way:
\begin{itemize}
\item {\bf INPUT:}
\begin{enumerate}
\item Superpotential $W(\{\phi_i\})$, a polynomial in variables $\phi_{i=1, \ldots, n}$.
\item Generators of GIOs: $r_j(\{\phi_i\})$, $j=1, \ldots, k$ polynomials in $\phi_i$.
\end{enumerate}

\item {\bf ALGORITHM:}
\begin{enumerate}
\item Define the polynomial ring $R = \IC[\phi_{i=1,\ldots,n}, y_{j=1,\ldots,k}]$.
\item Consider the ideal $I = \gen{\frac{\partial {W}}{\partial \phi_i}, y_j - r_j(\{\phi_i\})}$.
\item Eliminate all variables $\phi_i$ from $I \subset R$, giving the ideal $\cM$ in terms of $y_j$.
\end{enumerate}

\item {\bf OUTPUT:}\\
$\cM$ corresponds to the vacuum moduli space as an affine variety in $\IC[y_1, \ldots, y_k]$.
\end{itemize}

This paper focuses on discussing the output of this algorithm for the MSSM electroweak sector, considering various number of particle flavors.
The resulting affine varieties $\cM$ are intersections of homogeneous polynomials and, as such, we can write them as affine cones over a compact projective variety $\CB$ of one lower dimension. We will thus adopt the notation to which we have adhered for many years,
\bea \nn
\mbox{\fbox{$\cM=(k | d, \delta | m_1^{n_1} m_2^{n_2} \ldots )$}} &:=&
\mbox{Affine variety of complex dimension $d$, realized as an affine} \\ \nn
&&\mbox{cone over a projective variety of dimension $d-1$  and degree $\delta$,}\nn \\
&&\mbox{given as the intersection of $n_i$ polynomials of degree $m_i$ in $\IP^k$.}\nn \\  \label{eq:ac}
\eea

\section{Multi-generation Electroweak Models}\label{s:mutli}\setall

Presently, we contemplate a renewed effort to calculate the full geometry of the MSSM vacuum moduli space.
The computing power required for applying the previously described algorithm with $\sim 1000$ GIOs and $\sim 50$ fields is well beyond what is accessible by standard personal computers.
The use of supercomputers is envisaged.
For this reason, our goal here is significantly more modest, and we only unveil aspects of the geometry for the electroweak sector.
That is, we study a subsector of the full vacuum moduli space that is given by the additional constraints that the vacuum expectation values of the quark fields vanish:
\be
Q_{a, \alpha}^i=u_a^i=d_a^i=0 \ .
\ee
This is perhaps reasonable on phenomenological grounds as $SU(3)_C$ is an unbroken symmetry in Nature.
The non-vanishing GIOs that remain from the list in Appendix~\ref{GIOs} are noted in Table~\ref{gio-ew}.
\begin{table}[h]
{\begin{center}
\begin{tabular}{|c||c|c|c|}\hline
\mbox{Type} & \mbox{Explicit Sum} & \mbox{Index} & \mbox{Number} \\
\hline \hline
$LH$  & $L^i_\alpha H_\beta \eps$ & $i=1,2,\ldots N_{f}$ & $N_{f}$ \\ \hline
$H\barH$ & $H_\alpha \barH_\beta \eps$ & & 1  \\ \hline
$LLe$ & $L^i_\alpha L^j_\beta e^k \eps$ & $i,k=1,2,\ldots, N_{f};
j=1,\ldots,i-1$ & $N_{f}\cdot\binom{N_{f}}{2}$  \\ \hline
$L\barH e$ & $L^i_\alpha \barH_\beta \eps e^j$ & $i,j=1,2,\ldots, N_{f}$ & ${N_{f}}^2$ \\
\hline
\end{tabular}
\end{center}}{\caption{\label{gio-ew}{\sf Minimal generating set of the GIOs for the electroweak sector}.}}
\end{table}

The minimal renormalizable superpotential of the electroweak sector of the MSSM is then
\beq\label{minimalew}
W_{\rm minimal} = C^0 H_\alpha \barH_\beta \eps + \sum_{i,j} C^3_{ij} e^i L^j_{\alpha} \barH_\beta \eps \; .
\eeq
Henceforth, we are explicit about the sums on flavor indices $i,j$ but leave sums on $SU(2)_L$ indices $\alpha,\beta$ implicit.
The corresponding F-terms are
\bea
\frac{\partial W_{\rm minimal}}{\partial H_\alpha}&=& C^0 \barH_\beta \eps\; , \label{eq:nine} \\
\frac{\partial W_{\rm minimal}}{\partial \barH_\beta}&=& C^0 H_\alpha \eps + \sum_{i,j} C^3_{ij} e^i L^j_{\alpha} \eps\; ,\\
\frac{\partial W_{\rm minimal}}{\partial L^j_{\alpha}}&=& \sum_i C^3_{ij} e^i \barH_\beta \eps\; ,\\
\frac{\partial W_{\rm minimal}}{\partial e^i}&=& \sum_j C^3_{ij} L^j_{\alpha} \barH_\beta \eps\; . \label{eq:twelve}
\eea
In particular, this yields the following F-term equations for the Higgs fields:
\bea\label{barH0}
\barH_\beta=0 \; ,\\\label{minHb}
C^0 H_\alpha + \sum_{i,j}C^3_{ij}e^iL^j_{\alpha}=0 \; ,
\eea
from the $F_{H_\alpha}$ and $F_{\barH_\beta}$ terms, respectively.
The other two F-term equations (for the $e$ and $L$ fields) do not lead to extra constraints as the vanishing of $\barH_\beta$ renders them trivial.

In terms of the $\{r_i\}$, the only non-trivial GIOs that remain are the $LH$ and $LLe$ operators.
Indeed, $H\barH$ and $L\barH e$ vanishes by virtue of~\eqref{barH0}.
Furthermore,~\eqref{minHb} specifies the value of the $LH$ operators in terms of the $LLe$ operators.
Multiplying~(\ref{minHb}) by $L^i_\beta \eps$ and summing on $\alpha$ gives
\be
C^0 L^i_\alpha H_\beta \eps+\sum_{j,k}C^3_{jk}L^i_\alpha L^j_{\beta}e^k\eps=0\; .
\label{eq:shit1}
\ee
(We have taken $C^3_{ij} = C^3_{ji}.$)
Since there is a free index $i$ in~\eqref{eq:shit1}, there are $N_f$ linear equations which suppress the $LH$ variables as degrees of freedom in the vacuum moduli space.
Thus, only $LLe$ contributes to the dimension counting of the vacuum geometry and the moduli space reduces to an affine variety in $\IC[y_1, \ldots, y_k]$ with $k=1,\ldots, N_f\cdot\binom{N_f}{2}$ given by the relations among the $LLe$ polynomials.\footnote{By abuse of terminology, we identify the ring $\IC[y_1,\ldots,y_k]$ and its corresponding k-affine space $\IC^k$, whereby not making the distinction between the two, as is customary in the physics community.}
The remaining coordinates $y$ resulting from the $LH$ operators only provide an embedding into the bigger ring $\IC[y_1, \ldots, y_{k+N_f}]$.

\subsection{Relations, Syzygies, and Grassmannian}\label{secrel}

Having established that the moduli space is only given by the relations among the $LLe$ operators, let us study these relations\footnote{The relations are presented in~\cite{Gherghetta:1995dv}; however, here we add to that analysis by presenting and counting the redundancy of the relations (syzygies).} explicitly for some specified number of matter generations $N_f$.
Explicitly, $LLe = L^i_\alpha L^j_\beta e^k \eps$.
The flavor index $k$ of the electron can assume any of the $N_f$ possibilities.
The indices $i$ and $j$ must be different due to the contraction with the antisymmetric tensor.
There are therefore $\binom{N_f}{2}$ choices for the combination $L^i_\alpha L^j_\beta \eps$. 
Because the $SU(2)_L$ indices $\alpha,\beta$ only take values $1, 2$, there is an upper bound on the number of lepton doublets that we can introduce before certain composite operators perforce vanish.
Indeed, there are relations between combinations of $LLe$ operators.
Taking into account every possible index combination with $(i,j) \neq (m,n)$ and $k\neq p$, a bit of algebra allows us to deduce the relation
\be
(L^i_\alpha L^j_\beta e^k \eps)(L^m_\gamma L^n_\delta e^p \epsilon^{\gamma\delta})=(L^m_\alpha L^n_\beta e^k \eps)(L^i_\gamma L^j_\delta e^p \epsilon^{\gamma\delta}) ~.
\ee
These are the relations for the ideal, which we write as
\be
\langle \ (L^i_\alpha L^j_\beta e^k \eps)(L^m_\alpha L^n_\beta e^p \eps)-(L^m_\alpha L^n_\beta e^k \eps)(L^i_\alpha L^j_\beta e^p \eps)\ \rangle \; . \label{LLe}
\ee
In a slight abuse of convention, we have restricted the remit of sums over $SU(2)_L$ indices to lie within the parentheses when we write out operators with $LL$ fields explicitly. We will adopt this convention from now on.

Heuristically, given the form of the relations in the ideal, we can cast the defining relations as an equality of quotients:
\be\label{LLrel1}
\frac{L^i_\alpha L^j_\beta e^k \eps}{L^i_\alpha L^j_\beta e^p \eps}=\frac{L^m_\alpha L^n_\beta e^k \eps}{L^m_\alpha L^n_\beta e^p \eps} \; ,
\ee
where again the summation over $\alpha,\beta$ restricts to the numerator or the denominator.
The equality~\eqref{LLrel1} informs us that a set of operators with a common $e^k$ field will be linearly proportional to another set of operators with a common $e^p$ field ($k \neq p$).
In a strict mathematical sense,~\eqref{LLrel1} only applies when the operators are non-vanishing in order to avoid problems with divisions by zero.
Nevertheless, this notation is a convenient way to succinctly express the relations we have encountered, keeping in mind that the branches with vanishing operators must be taken into account as well.

To determine the dimension of the variety, we only need to count a minimal generating set of such equations.
There are $(N_f-1)\left(\binom{N_f}{2}-1\right)$ non-trivial constraints when $N_f\geq3$.

When $N_f \geq 4$, more relations occur from the $LL$ component of the $LLe$ operators.
We have
\be\label{LLrel2}
(L^i_\alpha L^j_\beta \eps)(L^k_\alpha L^l_\beta \eps)+(L^i_\alpha L^k_\beta \eps)(L^l_\alpha L^j_\beta \eps)+(L^i_\alpha L^l_\beta \eps)(L^j_\alpha L^k_\beta \eps)=0 \; .
\ee
The set of indices ${i,j,k,l}$ can be chosen, without loss of generality, to be in a strictly increasing order.
This implies that there are $\binom{N_f}{4}$ such relations.
Let us introduce
\bea
\nn
P^{ijkl}&:=&(L^i_\alpha L^j_\beta \eps)(L^k_\alpha L^l_\beta \eps)\; ,\\
P^{i(jkl)}&:=& \sum_{{\rm \; cyclic \; permutations \;}(jkl)}P^{ijkl} \; .
\eea
We can then readily write~(\ref{LLrel2}) in the compact form:
\be\label{LLrel}
P^{i(jkl)}=0\; .
\ee
In general, this set of equations will be highly redundant as {\em syzygies} (relations among the generators) begin to appear.
Among the polynomials $P$, we have the syzygies
\bea\label{LLSyz1} %
P^{i(jkl)}(L^i_\alpha L^m_\beta \eps)-P^{i(jkm)}(L^i_\alpha L^l_\beta \eps)+P^{i(jlm)}(L^i_\alpha L^k_\beta \eps)&=&P^{i(klm)}(L^i_\alpha L^j_\beta \eps)\; ,\\\label{LLSyz2}
P^{i(jkl)}(L^j_\alpha L^m_\beta \eps)-P^{i(jkm)}(L^j_\alpha L^l_\beta \eps)+P^{i(jlm)}(L^j_\alpha L^k_\beta \eps)&=&P^{j(klm)}(L^i_\alpha L^j_\beta \eps)\; .
\eea

These syzygies imply that the relations~(\ref{LLrel}) can be chosen such that the indices $i=1$ and $j=2$ without loss of generality.
Indeed, all other choices of indices are simply redundant equations.
To see this, we can use the first syzygy~(\ref{LLSyz1}) to generate all $P$s starting with an index $i=1$ from $P$s starting with an index $i=1$ and $j=2$.
Explicitly, with the conventions that indices are in a strictly increasing order, we observe that all $P^{1(klm)}$ for $k,l,m>2$ are given by the relation
\be
P^{1(2kl)}(L^1_\alpha L^m_\beta \eps)-P^{1(2km)}(L^1_\alpha L^l_\beta \eps)+P^{1(2lm)}(L^1_\alpha L^k_\beta \eps)=P^{1(klm)}(L^1_\alpha L^2_\beta \eps) \; .
\ee

Having established that we can generate every polynomial $P$ starting with an index $1$, we can use the second syzygy~(\ref{LLSyz2}) with a choice of index $i=1$ and any indices $m>l>k>j\geq3$ to show that every relation in~(\ref{LLrel}) with indices $(i,j,k,l)$ greater than $2$ are redundant:
\be
P^{1(jkl)}(L^j_\alpha L^m_\beta \eps)-P^{1(jkm)}(L^j_\alpha L^l_\beta \eps)+P^{1(jlm)}(L^j_\alpha L^k_\beta \eps)=P^{j(klm)}(L^1_\alpha L^j_\beta \eps).
\ee

We can now count the total number of independent relations:
\be
\#{\rm \ relations}= \binom{N_f-2}{2} \; .
\ee
This is due to the fact that the independent constraints in~\eqref{LLrel} are given by $P^{1(2kl)}=0$ only.
So we need to choose two indices ($k$ and $l$) among $N_f-2$ possibilities ($i=1$ and $j=2$ being fixed).

Taking all of the above counting together, the dimension of the vacuum moduli space will be
\be\label{dimformula}
N_f\cdot\binom{N_f}{2}-(N_f-1)\left(\binom{N_f}{2}-1\right)-\binom{N_f-2}{2} = 3N_f-4 \; .
\ee

The vacuum geometry can be understood as follows.
Explicitly, the index structure $LLe =  L^i_\alpha L^j_\beta e^k \eps$ shows that $L^i_\alpha L^j_\beta \eps$ furnishes, due to the antisymmetry, coordinates on the Grassmannian $Gr(N_f,2)$ of two-planes in $\IC^{N_f}$.
The freely indexed $e^k$, on the other hand, gives simply a copy of $\IP^{N_f-1}$. Topologically, the geometry is then given by (the affine cone over) $Gr(N_f,2) \times \IP^{N_f-1}$.

In fact, the above dimension counting simply corresponds to the dimension of the Grassmannian
\be
\dim Gr(n,r) = r(n-r) \; ,
\ee
given by the $LL$ part of the operators. Therefore, according to the product $Gr(N_f,2) \times \IP^{N_f-1}$, the affine dimension is obtained
%
\begin{equation}
\dim \cM_{EW} = 2(N_f - 2) + N_f = 3N_f - 4 \ .
\end{equation}
Thus, the dimension always increases by three when we add another generation of matter fields to the electroweak sector.

It is a remarkable fact that the dimension increases by the same increment as the number of fields, despite the number of GIOs growing much faster.
\begin{table}[h]
{\begin{center}
\begin{tabular}{|c||c|c|c|c|c|c|c|}\hline
$N_f$ & 1 & 2 & 3 & 4 & 5 & 6 & $\ldots$ \\ \hline\hline
number of fields & 5 & 8 & 11 & 14 & 17 & 21 & $\ldots$ \\ \hline
number of $LLe$ generators & 0 & 2 & 9 & 24 & 50 & 90 & $\ldots$ \\ \hline
vacuum dimension & 0 & 2 & 5 & 8 & 11 & 14 & $\ldots$ \\ \hline
\end{tabular}
\end{center}}{\caption{\label{vac-dim}{\sf Vacuum geometry dimension according to the number generations $N_f$}.}}
\end{table}

In the following subsections, we will study in greater detail the geometry for the cases $N_f = 2,3, 4, 5$.

\subsection{Counting Operators with Hilbert Series}\label{secminideal}

The Hilbert series provides technology for enumerating GIOs in a supersymmetric quantum field theory.
For a variety ${\cal M} \subset \IC[y_1,\ldots,y_k]$, the Hilbert series supplies a generating function:
\be
H(t) = \sum_{n=-\infty}^\infty \mathrm{dim}\, {\cal M}_n \, t^n = \frac{P(t)}{(1-t)^d} ~.
\ee
This has a geometrical interpretation.
The quantity $\mathrm{dim}\, {\cal M}_n$ that appears in the sum constitutes the (complex) dimension of the graded pieces of ${\cal M}$.
That is to say, it represents the number of independent polynomials of degree $n$ on ${\cal M}$.
When we write the Hilbert series as a ratio of polynomials, the numerator and the denominator both have integer coefficients.
The dimension of the moduli space is $d$, which is the order of the pole at $t=1$.
Via the plethystic exponential and the plethystic logarithm, the Hilbert series encodes information about the chiral ring and geometric features of the singularity from which the supersymmetric gauge theory under consideration arises.
It should be emphasized that the Hilbert series is not a topological invariant and can be represented in many ways.
For our purposes, an important caveat is that Hilbert series depends on the embedding of the variety within the polynomial ring~\cite{m2book}.
The reader is referred to~\cite{Benvenuti:2006qr} for an account of the importance of the Hilbert series in the context of gauge theories.

In this investigation, we will write the Hilbert series for the vacuum moduli space of the electroweak sector for various values of $N_f$.
The Hilbert series is a mathematical object that can be constructed using standard techniques in computational algebraic geometry.
Knowledge of certain properties of the Hilbert series will allow us to characterize the structure of the vacuum geometry.

\subsection{Vacuum Geometry}

Let us introduce the following label for the non-vanishing $LLe$ operators
\be\label{ynotation}
y_{I+C(N_f,2)\cdot (k-1)}=L^i_\alpha L^j_\beta e^k \eps\;,
\ee
where $C(N_f,2)=\binom{N_f}{2}$ are binomial coefficients and $I=1,\ldots,C(N_f,2)$ accounts for the $(i,j)$ index combinations from $L^i_\alpha L^j_\beta  \eps$.
With this notation, the first set of relations~(\ref{LLrel1}) becomes,
\be\label{yindexnotation}
\frac{y_{I+C(N_f,2)\cdot (k-1)}}{y_{I+C(N_f,2)\cdot (l-1)}}=\frac{y_{J+C(N_f,2)\cdot (k-1)}}{y_{J+C(N_f,2)\cdot (l-1)}}\; ,
\ee
for $I,J=1,\ldots,C(N_f,2)$ and $k,l=1,\ldots,N_f$.
For a minimal set of equations, we can then choose, for instance, $I<J$ and $k=1$.
Moreover, there will be a set of relations from~(\ref{LLrel}) among each $y_{I+C(N_f,2)\cdot (k-1)}$ for a fixed $k$ when $N_f$ is large enough ($N_f\geq 4$).
We cannot write these relations explicitly for all of the $N_f$ at once, so let us consider each value of $N_f$ separately.\footnote{
We stopped this investigation at $N_f=5$ due to limitations of computer power.}

\subsubsection*{$N_f=2$}

In this case, we have only two $LLe$ operators for the two $e^i$ fields. Thus, we cannot have any relations and the vacuum moduli space is trivially the plane $\cM=\IC^2$.

\subsubsection*{$N_f=3$}
We have nine $LLe$ operators and the vacuum moduli space will be an algebraic variety in $\IC^9$.
With the above notation~\eqref{ynotation}, the relations~\eqref{yindexnotation} become
\bea\label{relfrac1}
\frac{y_1}{y_4}=\frac{y_2}{y_5}=\frac{y_3}{y_6} &,& \\\label{relfrac2}
\frac{y_1}{y_7}=\frac{y_2}{y_8}=\frac{y_3}{y_9} &.&
\eea
This leads to an ideal given by nine quadratic polynomials
\bea\nn
 \langle \ y_1y_5-y_2y_4,\; y_1y_6-y_3y_4,\; y_2y_6-y_3y_5,\\
\label{EW3gen}
y_1y_8-y_2y_7,\; y_1y_9-y_3y_7,\; y_2y_9-y_3y_8,\\ \nn y_4y_8-y_5y_7,\; y_4y_9-y_6y_7,\; y_5y_9-y_6y_8 \ \rangle \ . 
\eea

We count $(3-1)\left(\binom{3}{2}-1\right)=4$ equalities in~\eqref{relfrac1} and~\eqref{relfrac2} and we find that the resulting moduli space $\cM$ is an irreducible %
five-dimensional affine variety given by an affine cone over a base manifold $\CB$ of dimension four.
As a projective variety, $\CB$ has degree six and is described by the (non-complete) intersection of nine quadratics in $\IP^8$, which agrees with the results of~\cite{Gray:2006jb,He:2014loa}.
This can be summarized according to the standard notation~\eqref{eq:ac} as
\begin{equation}
\cM_{\rm EW} = (8|5,6|2^9)\; . \label{M_EW}
\end{equation}

The variety $\cM$ is in fact a non-compact toric Calabi--Yau.
The reader is referred to Appendix~\ref{ap:toricCY} for a detailed discussion on toric affine Calabi--Yau spaces.
Indeed, its Hilbert series is given by
\be\label{H_EW}
\frac{1+4t+t^2}{(1-t)^5} \ ,
\ee
and is {\em palindromic}.
By this, we mean simply that the numerator of the Hilbert series can be written in the form
\begin{equation} P(t) = \sum_{k=0}^{N} a_k t^k \, , \label{palindromic} \end{equation}
with the simple property that $a_k = a_{N-k}$.
It has been shown~\cite{Stanley} that the numerator of the Hilbert series of a graded Cohen--Macaulay domain $X$ is palindromic if and only if $X$ is Gorenstein\footnote{
In this work, for all the varieties considered, $\cM$ is always an integral domain
arithmetically Cohen--Macaulay. Hence, we will loosely use the correspondence that, for affine varieties, palindromic Hilbert series means Calabi--Yau.
}.
For affine varieties, the Gorenstein property implies that the geometry is Calabi--Yau.
Additional discussion of this point can be found in~\cite{Gray:2008yu}, and in Appendix~\ref{ap:toricCY} for clarifications on the Gorenstein property.
The vacuum moduli spaces we obtain are non-compact.

As mentioned in Section~\ref{secrel}, the topology is given by the cone over $Gr(3,2) \times \IP^{2}$.
The Grassmannian $Gr(3,2)$ is exactly $\IP^2$, while the second $\IP^{2}$ comes from $\IP^{N_f-1}$.
Therefore, the affine five-dimensional vacuum space described by~\eqref{M_EW} is none other than the cone over the Segr\`e embedding of $\IP^2 \times \IP^2$ into $\IP^8$.
We remind the reader that this is the following space.
Take $[x_0:x_1:x_2]$ and $[z_0:z_1:z_2]$ as the homogeneous coordinates on the two $\IP^2$s respectively and consider the quadratic map
\begin{equation} \label{Segre}
\begin{array}{ccccc}
\IP^2 & \times & \IP^2 & \longrightarrow & \IP^8 \\
\mbox{$[x_0 : x_1 : x_2]$}
& &  
\mbox{$[z_0 : z_1 : z_2]$}
& \rightarrow & x_i z_j \\
\end{array} ~,
\end{equation}
where $i,j=0,1,2$ give precisely the $3^2 = 9$ homogeneous coordinates of $\IP^8$.
Explicitly, upon elimination, this is exactly the nine quadrics with the Hilbert series as given in~\eqref{M_EW} and~\eqref{H_EW}.
We also point out that this Segr\`e variety is the only Severi variety of projective dimension four. Later we will re-encounter Severi varieties of a unique nature in dimension two.

For the reader's convenience, let us recall the definition of a Severi variety~\cite{severi,zak,ab}.\footnote{
We are grateful to Sheldon Katz for his insight and for mentioning Severi varieties to us.}
It is a classic result of Hartshorne--Zak~\cite{zak} that any smooth non-degenerate algebraic variety $X$ of (complex) dimension $n$ embedded into $\IP^m$ with $m < \frac32n + 2$ has the property that its secant variety $Sec(X)$ --- \textit{i.e.}, the union of all the secant and tangent lines to $X$ --- is equal to $\IP^m$ itself.
The limiting case\footnote{
In general, a $k$-Scorza variety is a smooth projective variety, of maximal dimension such that its $k-1$ secant variety is not the whole of the ambient projective space. The Severi variety is the case of $k=2$.
} of $m = \frac32 n + 2$ and $Sec(X) \ne \IP^n$ is called a {\bf Severi variety}.
The classification theorem of Zak~\cite{zak} states that there are only four Severi varieties (the dimensions are precisely equal to $2^q$ with $q$ the dimension of the four division algebras):
\begin{enumerate}
\item[] $n=2$: The Veronese surface $\IP^2 \hookrightarrow \IP^5$;
\item[] $n=4$: The Segr\`e variety $\IP^2 \times \IP^2 \hookrightarrow \IP^8$;
\item[] $n=8$: The Grassmannian $Gr(6,2)$ of two-planes in $\IC^6$, embedded into $\IP^{14}$;
\item[] $n=16$: The Cartan variety of the orbit of the highest weight vector of a certain non-trivial representation of $E_6$.
\end{enumerate}
Of these, only two are isomorphic to (a product of) projective space, namely $n=2,4$.
Remarkably, these are the two that show up as the vacuum geometry of the electroweak sector when $N_f=3$.

The connection with Severi varieties could be profound.
Indeed, it was discussed in~\cite{ab} that these four spaces are fundamental to mathematics in the following way. It is well-known that there are four division algebras: the real numbers $\IR$, the complex numbers $\IC$, the quaternions $\IH$, and the octonions $\IO$, of, respectively, real dimension $1,2,4,8$. Consider the projective planes formed out of them, {\it viz.}, $\IR\IP^2$, $\IC\IP^2$, $\IH\IP^2$ and $\IO\IP^2$, of real dimension $2,4,8,16$. We have, of course, encountered $\IC\IP^2$ repeatedly in our above discussions.
The complexification of these four spaces, of complex dimension $2,4,8,16$ are precisely homeomorphic to the four Severi varieties.
Amazingly, they are also homogeneous spaces, being quotients of Lie groups.
In summary, we can tabulate the four Severi varieties
\begin{equation}\label{t:severi}
\begin{array}{|c|c|c|}\hline
\mbox{Projective Planes} & \mbox{Severi Varieties} & \mbox{Homogeneous Spaces}
\\ \hline  \hline
\IR\IP^2 & \IC\IP^2 & SU(3) / S( \ U(1) \times U(2) \ ) \\ \hline 
\IC\IP^2 & \IC\IP^2 \times \IC\IP^2 & 
   SU(3)^2  / S( \ U(1) \times U(2) \ )^2 \\ \hline 
\IH\IP^2 & Gr(6,2) &  SU(6) / S( \ U(2) \times U(4) \ ) \\ \hline 
\IO\IP^2 & S & E_6 / Spin(10)U(1) 
\\ \hline
\end{array}
\end{equation}

Returning to our present case of $n=4$, the embedding~\eqref{Segre} can be understood in terms of the previously defined $y$ variables. Let us consider the following change of variables,
\be\label{newvar3}
\ba{lll}
y_1 \rightarrow z_0x_2 ~, & y_2 \rightarrow z_0x_1 ~, & y_3 \rightarrow z_0x_0 ~, \\
y_4 \rightarrow z_1x_2 ~, & y_5 \rightarrow z_1x_1 ~, & y_6 \rightarrow z_1x_0 ~, \\
y_7 \rightarrow z_2x_2 ~, & y_8 \rightarrow z_2x_1 ~, & y_9 \rightarrow z_2x_0 ~.
\ea
\ee
The $z$ coordinates labels the $\IC^3$ due to the $e$ fields, while the $x$ coordinates label the Grassmannian due to $LL$. With these variables, the relations~\eqref{relfrac1} and~\eqref{relfrac2} are automatically satisfied.

This variety is also toric, as can be seen by the binomial nature of the polynomial ideal~\eqref{EW3gen}. Its toric diagram can be presented as follows,
\begin{equation}
\kN =
{\scriptsize
\left(
\begin{array}{ccccc}
 1 & 0 & 1 & 0 & 0 \\
 1 & 1 & 0 & 0 & 0 \\
 1 & 1 & 0 & 1 & 0 \\
 0 & 0 & 1 & 0 & 0 \\
 0 & 1 & 0 & 0 & 0 \\
 0 & 1 & 0 & 1 & 0 \\
 0 & -1 & 1 & -1 & 1 \\
 0 & 0 & 0 & -1 & 1 \\
 0 & 0 & 0 & 0 & 1 \\
\end{array}
\right)
} \ ,
\end{equation}
where each row of the matrix corresponds to the vectors generating the toric cone. They are five-dimensional vectors as required for a five-dimensional variety. We have nine of them, as expected from the nine quadratics in~\eqref{EW3gen}. Further details on the toric diagrams are given in Appendix~\ref{toric}.

Finally, we found that the Hodge diamond of the base space $\CB$ is given by
\begin{equation}
h^{p,q}(\CB) \; = \;
{\begin{array}{ccccccccc}
&&&&h^{0,0}&&& \\
&&&h^{0,1}&&h^{0,1}&&& \\
&&h^{0,2}&&h^{1,1}&&h^{0,2}&& \\
&h^{0,3}&&h^{1,2}&&h^{1,2}&&h^{0,3}& \\
h^{0,4}&&h^{1,3}&&h^{2,2}&&h^{1,3}&&h^{0,4} \\
&h^{0,3}&&h^{1,2}&&h^{1,2}&&h^{0,3}& \\
&&h^{0,2}&&h^{1,1}&&h^{0,2}&& \\
&&&h^{0,1}&&h^{0,1}&&& \\
&&&&h^{0,0}&&&& \\
\end{array}}
\; = \;
{\begin{array}{ccccccccc}
&&&& 1 &&&& \\
&&& 0 && 0 &&& \\
&& 0 && 2 && 0 && \\
& 0 && 0 && 0 && 0 & \\
0 && 0 && 3 && 0 && 0  \\
& 0 && 0 && 0 && 0 & \\
&& 0 && 2 && 0 && \\
&&& 0 && 0 &&& \\
&&&& 1 &&&& \\
\end{array}}\, ,
\label{hodge2}
\end{equation}
which has the peculiar property to be non-vanishing in its diagonal only. This Hodge diamond is consistent with the Hodge diamond of $\IP^2 \times \IP^2$ as can be seen by using the K\"unneth formula. Of course, as is with a later example in \eqref{hodge3}, having the same Hodge diamond is only a statement of topology, our analysis is more refined in that we can identify what the variety actually is.

We note that the surface itself was first identified in~\cite{Gray:2006jb}, but the only information that could be gleaned about the manifold at that time is that encapsulated by the notation of~(\ref{M_EW}).\footnote{However, a typographical error in~\cite{Gray:2006jb} presented the variety as given by six quadratics.}
Since that time, improvements in computing and software have allowed us to calculate both the Hilbert series and the above Hodge diamond, as well as leading to a complete understanding of its geometrical nature.

\subsubsection*{$N_f=4$}
For the case of four flavors, we have twenty-four $LLe$ operators.
The first set of constraints~\eqref{yindexnotation} leads to $(4-1)\left(\binom{4}{2}-1\right)=15$ equations.
With the notation defined in~\eqref{yindexnotation} and the simplified choice of $LL$ labeling, we have
\bea\label{rel4gen}
\frac{y_1}{y_7}=\frac{y_2}{y_8}=\frac{y_3}{y_9}=\frac{y_4}{y_{10}}=\frac{y_5}{y_{11}}=\frac{y_6}{y_{12}} &,&\\
\frac{y_1}{y_{13}}=\frac{y_2}{y_{14}}=\frac{y_3}{y_{15}}=\frac{y_4}{y_{16}}=\frac{y_5}{y_{17}}=\frac{y_6}{y_{18}} &,&\\\label{rel4genend}
\frac{y_1}{y_{19}}=\frac{y_2}{y_{20}}=\frac{y_3}{y_{21}}=\frac{y_4}{y_{22}}=\frac{y_5}{y_{23}}=\frac{y_6}{y_{24}} &.&
\eea
Moreover, we need to take care of the constraints from the second set of relations~\eqref{LLrel}.
Due to the fact that we only have six possible $LL$ operators, we will have only one relation among them, given by 
\begin{equation} \label{plucker}
(L^1_\alpha L^2_\beta \eps)(L^3_\alpha L^4_\beta \eps)-(L^1_\alpha L^3_\beta \eps)(L^2_\alpha L^4_\beta \eps)+(L^1_\alpha L^4_\beta \eps)(L^2_\alpha L^3_\beta \eps)=0\, . \end{equation}
We can multiply this relation with any $e$ field to obtain the relations among $LLe$ operators, translated into the $y$ variable.
For $e^1$, we have,
\bea\label{rel4gen2}
y_1y_6+y_3y_4-y_2y_5=0 \;,
\eea
while for the other three $e$ fields, we have,
\bea
\nn
y_7y_{12}+y_9y_{10}-y_8y_{11}=0 \; ,\\
\nn
y_{13}y_{18}+y_{15}y_{16}-y_{14}y_{17}=0\; ,\\
y_{18}y_{24}+y_{21}y_{22}-y_{20}y_{23}=0\; .
\eea

It is straightforward to see that these equations do not lead to additional constraints.
Indeed, using~(\ref{rel4gen})--(\ref{rel4genend}) we can easily recover these from~(\ref{rel4gen2}).
Therefore, we have in total $\binom{4-2}{2}=1$ relation as expected.
The dimension of the vacuum moduli space is therefore $24-15-1=8$.
In fact, we have an irreducible eight-dimensional algebraic variety given by
\be
\cM = (23|8,70|2^{100}).
\ee

In general, for $N_f > 3$, the Grassmannian does not degenerate to projective space though the geometry still corresponds to some embedding of $Gr(N_f,2) \times \IP^{N_f-1}$ into higher dimensional space. The relation~\eqref{plucker} is none other than the Pl\"ucker relation for the Grassmannian $Gr(4,2)$.
Geometrically, however, there are no special names for birational embedding of products of Grassmannians with projective space, as was with the Segr\`e case.

The embedding could again be understood from a change of variables. Writing the Pl\"ucker coordinate for $Gr(4,2)$ as $[x_0:x_1:x_2:x_3:x_4:x_5]$ and taking $[z_0:z_1:z_2:z_3]$ for $\IP^3$, we can consider the change of variables,
\be\label{newvar4}
\ba{llllll}
y_1 \rightarrow z_0x_0 ~, & y_2 \rightarrow z_0x_1 ~, & y_3 \rightarrow z_0x_2 ~, & y_4 \rightarrow z_0x_3 ~, & y_5 \rightarrow z_0x_4 ~, & y_6 \rightarrow z_0x_5 ~,\\
y_7 \rightarrow z_1x_0 ~, & y_8 \rightarrow z_1x_1 ~, & y_9 \rightarrow z_1x_2 ~, & y_{10} \rightarrow z_1x_3 ~, & y_{11} \rightarrow z_1x_4 ~, & y_{12} \rightarrow z_1x_5 ~,\\
y_{13} \rightarrow z_2x_0 ~, & y_{14} \rightarrow z_2x_1 ~, & y_{15} \rightarrow z_2x_2 ~, & y_{16} \rightarrow z_2x_3 ~, & y_{17} \rightarrow z_2x_4 ~, & y_{18} \rightarrow z_2x_5 ~,\\
y_{19} \rightarrow z_3x_0 ~, & y_{20} \rightarrow z_3x_1 ~, & y_{21} \rightarrow z_3x_2 ~, & y_{22} \rightarrow z_3x_3 ~, & y_{23} \rightarrow z_3x_4 ~, & y_{24} \rightarrow z_3x_5 ~.
\ea
\ee
Imposing the Pl\"ucker relation
\be\label{Pluck}
x_0x_5-x_1x_4+x_2x_3=0 \ ,
\ee
for the coordinates $[x_0:x_1:x_2:x_3:x_4:x_5]$, all required relations are then satisfied.

Using algebraic geometry packages~\cite{mac,sing}, we obtain the Hilbert series
\be
\frac{1+16t+36t^2+16t^3+t^4}{(1-t)^8} \; .
\ee
Noting the palindromic property of the numerator, again we have an affine Calabi--Yau geometry.
However, the additional condition~\eqref{rel4gen2} is not explicitly toric.
We have this remarkable fact that only three generations of particles will provide explicitly toric geometries.
Indeed, any number above three will have relations such as the one above.

\subsubsection*{$ N_f=5$}

To illustrate the syzygies, let us look at the case with $N_f=5$.
There are $50$ $LLe$ operators.
The relations~\eqref{LLrel1} therefore contain $36$ equalities:
\bea %
\nn
\frac{y_1}{y_{11}}=\frac{y_2}{y_{12}}=\frac{y_3}{y_{13}}=\frac{y_4}{y_{14}}=\frac{y_5}{y_{15}}=\frac{y_6}{y_{16}}=\frac{y_6}{y_{17}}=\frac{y_7}{y_{18}}=\frac{y_8}{y_{19}}=\frac{y_{10}}{y_{20}} &,&\\
\nn
\frac{y_1}{y_{21}}=\frac{y_2}{y_{22}}=\frac{y_3}{y_{23}}=\frac{y_4}{y_{24}}=\frac{y_5}{y_{25}}=\frac{y_6}{y_{26}}=\frac{y_6}{y_{27}}=\frac{y_7}{y_{28}}=\frac{y_8}{y_{29}}=\frac{y_{10}}{y_{30}} &,&\\
\nn
\frac{y_1}{y_{31}}=\frac{y_2}{y_{32}}=\frac{y_3}{y_{33}}=\frac{y_4}{y_{34}}=\frac{y_5}{y_{35}}=\frac{y_6}{y_{36}}=\frac{y_6}{y_{37}}=\frac{y_7}{y_{38}}=\frac{y_8}{y_{39}}=\frac{y_{10}}{y_{40}} &,&\\\label{rel5genend}
\frac{y_1}{y_{41}}=\frac{y_2}{y_{42}}=\frac{y_3}{y_{43}}=\frac{y_4}{y_{44}}=\frac{y_5}{y_{45}}=\frac{y_6}{y_{46}}=\frac{y_6}{y_{47}}=\frac{y_7}{y_{48}}=\frac{y_8}{y_{49}}=\frac{y_{10}}{y_{50}} &.&
\eea

The relations obtained from~\eqref{LLrel} lead to the $\binom{5}{4}=5$ equations:
\bea\label{1refNf5}
y_1y_6+y_3y_4-y_2y_5=0 \; ,\\\label{2refNf5}
y_1y_9+y_3y_7-y_2y_8=0\; ,\\\label{3refNf5}
y_1y_{10}+y_5y_7-y_4y_8=0\; ,\\\label{4refNf5}
y_2y_{10}+y_6y_7-y_4y_9=0\; ,\\\label{5refNf5}
y_3y_{10}+y_6y_8-y_5y_9=0\; .
\eea
The other relations for $e^i$ with $i\neq 1$ will not bring additional constraints due to the first set of relations~\eqref{LLrel1}.
Now, we claimed that only the $\binom{5-2}{2}=3$ relations from
$
P^{1(2kl)}=0
$
are relevant.
Indeed, we can multiply~\eqref{1refNf5},~\eqref{2refNf5}, and~\eqref{3refNf5} by the appropriate $y$ variable to obtain
\bea
y_2(y_1y_{10}+y_5y_7-y_4y_8)+(y_1y_6+y_3y_4-y_2y_5)y_7-y_4(y_1y_9+y_3y_7-y_2y_8)= \nn \\
y_1(y_2y_{10}+y_6y_7-y_4y_9)
=0\;
\eea
yielding~\eqref{4refNf5} and
\bea
y_3(y_1y_{10}+y_5y_7-y_4y_8)+(y_1y_6+y_3y_4-y_2y_5)y_8-y_5(y_1y_9+y_3y_7-y_2y_8) = \nn \\
y_1(y_3y_{10}+y_6y_8-y_5y_9)
=0\;
\eea
yielding~\eqref{5refNf5}.
Thus we only have three genuine relations and the dimension of the space is $50-36-3=11$ as expected from~\eqref{dimformula}.

Using algebraic geometry packages~\cite{mac,sing}, we found an irreducible algebraic variety given by
\be
\cM = (49|11,1050|2^{525}).
\ee
Its Hilbert series is
\be
\frac{1+39t+255t^2+460t^3+255t^4+39t^5+t^6}{(1-t)^{11}} \; ,
\ee
and again, we have an affine Calabi--Yau space which is not itself an explicit toric variety.

\section{Multiple Higgs Generations}\label{sec:four}\setall

Before considering the vacuum geometry of the MSSM electroweak sector in the presence of neutrinos, let us first stop to consider what effect changing the number of generations of Higgs multiplets might have on the results we have already obtained.
Let $N_h$ denote the number of pairs of Higgs doublets in the theory, and let us restrict ourselves to the case in which $N_h \leq N_f$.
Both the up-type Higgs doublet $H_{\alpha}^k$ and down-type Higgs doublet $\barH_{\alpha}^k$ must now be labeled by a generation index $k=1,\ldots,N_h$.
The Yukawa coupling matrix $C^3$ is now promoted to a three-index tensor $C^3_{ij,k}$, where $i,j=1,\ldots,N_f$ and $k=1,\ldots,N_h$, and we imagine a bilinear term that allows arbitrary mixing among the Higgs generations: $C^0_{ij}$, with indices $i,j=1,\ldots,N_h$. The range of the indices should be clear from the context and, from now on, we will leave the range implicit.
The GIOs here are summarized in Table~\ref{gio-ew-Higgs}.

\begin{table}[h!!!]
{\begin{center}
\begin{tabular}{|c||c|c|c|}\hline
\mbox{Type} & \mbox{Explicit Sum} & \mbox{Index} & \mbox{Number} \\
\hline \hline
$LH$  & $L^i_\alpha H^j_{\beta} \eps$ & $i=1,\ldots, N_{f}$; $j=1,\ldots, N_{h}$ & $N_{f}\cdot N_h$ \\ \hline
$H\barH$ & $H^i_{\alpha} \barH^j_{\beta} \eps$ & $i,j=1,\ldots, N_{h}$ & $N_h^2$  \\ \hline
$LLe$ & $L^i_\alpha L^j_\beta e^k \eps$ & $i,k=1, \ldots, N_{f};
j=1,\ldots,i-1$ & $N_{f}\cdot\binom{N_{f}}{2}$  \\ \hline
$\barH \barH e$ & $\barH^i_\alpha \barH^j_\beta e^k \eps$ & $i=1,\ldots, N_{h}$;
$j=1,\ldots,i-1$; $k=1,\ldots, N_{f}$ & $N_{f}\cdot\binom{N_{h}}{2}$  \\ \hline
$L\barH e$ & $L^i_\alpha \barH^k_{\beta} \eps e^j$ & $i,j=1,\ldots, N_{f}$; $k=1,\ldots, N_{h}$ & ${N_{f}}^2\cdot N_h$ \\
\hline
\end{tabular}
\end{center}}{\caption{\label{gio-ew-Higgs}{\sf Minimal generating set of the GIOs for the electroweak sector, for number of Higgs doublets $N_h >1$}.}}
\end{table}

Clearly, such a construction would immediately engender phenomenological objections to the likely large flavor changing neutral current processes such a model would permit ({\em e.g.}, large rates for $\mu \to e \gamma$ processes, etc.).
But our interest here is to ask whether such a model, {\em a priori} possible, or even natural, from the point of view of an underlying string theory,\footnote{
For example, in the spirit of trinification, there are three generations of Higgs doublets in the $\Delta_{27}$ model~\cite{aiqu,bjl}, which embeds the Standard Model on the worldvolume of a single D$3$-brane. Three generations of Higgses are also expected in models based on $E_6$ gauge groups~\cite{Hewett:1988xc,King:2005jy,Kang:2007ib}.
} has a geometry that is significantly different from that which arises in the one generation case.

When $N_h \neq 1$, we expect a larger set of GIOs and thus, at least na\"{\i}vely, we might expect the vacuum moduli space to be of larger dimension than the $N_h=1$ case.
Indeed, the operator types $LH$ and $L\barH e$ from Table~\ref{gio-ew} now represent $N_f \cdot N_h$ objects, while the bilinear $H\barH$ now represents $N_h^2$ terms.
Since the lepton doublet $L$ and the down-type Higgs $\barH$ have the same $SU(2)_L\times U(1)_Y$ quantum numbers, we can extend the list of GIOs in a straightforward manner.
A new operator type in the electroweak sector is $\barH \barH e$.
It is the analog of the $LLe$ term and, because of the implicit antisymmetric tensor, is allowed only in the case of multiple Higgs doublets.

The minimal superpotential we consider in this section is therefore
\beq\label{minimalewHiggs}
W_{\rm minimal} = \sum_{i,j} C^0_{ij} H^i_{\alpha} \barH^j_{\beta} \eps + \sum_{i,j,k} C^3_{ij,k} e^i L^j_{\alpha} \barH^k_{\beta} \eps \; .
\eeq
The F-terms are modified from those of~\eqref{eq:nine}--\eqref{eq:twelve} to read
\bea
\frac{\partial W_{\rm minimal}}{\partial H^i_{\alpha}}&=& \sum_j C^0_{ij} \barH^{j}_{\beta} \eps\; ,\\
\frac{\partial W_{\rm minimal}}{\partial \barH^{k}_{\beta}}&=& \sum_i C^0_{ik} H^i_{\alpha} \eps + \sum_{i,j} C^3_{ij,k} e^i L^j_{\alpha} \eps\; ,\\
\frac{\partial W_{\rm minimal}}{\partial L^j_{\alpha}}&=& \sum_{i,k} C^3_{ij,k} e^i \barH^{k}_{\beta} \eps\; ,\\
\frac{\partial W_{\rm minimal}}{\partial e^i}&=& \sum_{j,k} C^3_{ij,k} L^j_{\alpha} \barH^k_{\beta} \eps\; .
\eea
This leads to the following F-term equations for the Higgs fields:
\bea\label{barH0temp}
\barH^j_\beta=0 \; ,\\
\label{minHbtemp}
\sum_i C^0_{ik} H^i_{\alpha} + \sum_{i,j}C^3_{ij,k}e^iL^j_{\alpha}=0 \; .
\eea
Once again, the vanishing of $\barH^j_{\beta}$ leaves the other two F-term equations trivially satisfied. Note that~(\ref{minHbtemp}) now represents $N_h$ separate constraint equations, labeled by the free index $k$.

As before, the necessary vanishing of the $N_h$ fields $\barH^j_{\beta}$ in the vacuum ensures the vanishing of $H\barH$, $L\barH e$, and the new operators $\barH \barH e$ in the vacuum.
Similarly, we have a set of relations for the $LH$ operators formed by contraction of~(\ref{minHbtemp}) with $L^l_\beta \eps$. We obtain,
\be
C^0_{ik} H^i_\alpha L^l_\beta \eps+\sum_{i,j}C^3_{ij,k} e^i L^j_\alpha L^l_{\beta} \eps=0\; .
\ee
These are $N_f\cdot N_h$ linear equations (from the $k$ and $l$ free indices) which suppress the $LH$ variables as degrees of freedom in the vacuum moduli space.
Thus, once again, only $LLe$ contributes to the dimension counting of the vacuum geometry and the moduli space reduces to an affine variety in $\IC[y_1, \ldots, y_k]$ with $k=1,\ldots, N_f\cdot\binom{N_f}{2}$ given by the relations among the $LLe$ polynomials.
Therefore, we see that the vacuum moduli space is completely independent of the number of Higgs generations in the model.
From this analysis, we can argue that the vacuum moduli space of the MSSM electroweak sector, without neutrinos, will be a non-compact Calabi--Yau for all values of $N_h \le N_f$.
Despite our na\"{\i}ve intuition, the dimension of the vacuum geometry is unchanged, though the addition of extra Higgs degrees of freedom allows for an embedding into the now larger polynomial ring $\IC[y_1, \ldots, y_{k+N_f\cdot N_h}]$.

\section{Right-handed Neutrinos}\label{sec:five}\setall

Neutrinos have mass.
The mass can be generated by a coupling of a right-handed neutrino $\nu$ to the up-type Higgs field.
Because the neutrino carries no charge under $SU(3)_C\times SU(2)_L\times U(1)_Y$, this means that the field $\nu$ is itself a GIO.
On physical grounds, $\nu$ should not appear by itself in a phenomenological superpotential as this would be a tadpole, which we can remove by a field redefinition.
We may have composite operators involving $\nu$, the simplest being $\nu^i \nu^j$, which introduces Majorana mass terms to the Lagrangian.

When considering right-handed neutrino fields, we previously noticed that some of the dimensions of the vacuum moduli space geometry get lifted.
The resulting geometry becomes a three-dimensional Veronese surface for the case of three generations of particles~\cite{Gray:2005sr,Gray:2006jb,He:2014loa}.
In this paper, we would like to consider cases with different number of particle families and understand the role of the GIOs for the structure of the vacuum geometry, focusing on the Majorana mass terms.

Let us consider the electroweak sector as described in the previous section with the addition of extra right-handed neutrinos fields as presented in Table~\ref{rhneut}.
\begin{table}[h]
{\begin{center}
\begin{tabular}{|c||c|c|c|}\hline
\mbox{Type} & \mbox{Explicit Sum} & \mbox{Index} & \mbox{Number} \\
\hline \hline
$\nu$  & $\nu^i$ & $i=1,2,\ldots, N_{f}$ & $N_{f}$ \\ \hline
\end{tabular}
\end{center}}{\caption{\label{rhneut}{\sf Right-handed neutrinos fields}.}}
\end{table}

The corresponding superpotential terms are given by
\be
W_{\rm neutrinos} = \sum_{i,j}C^4_{ij} \nu^i\nu^j+\sum_{i,j} C^5_{ij} \nu^i L^j_{\alpha} H_\beta \eps ~.
\ee
Again, these interactions respect R-parity.
Taking derivatives of the superpotential yields the F-terms:
\bea
\frac{\partial W_{\rm neutrinos}}{\partial H_\beta}&=&\sum_{i,j} C^5_{ij} \nu^i L^j_{\alpha}\eps ~,\\
\frac{\partial W_{\rm neutrinos}}{\partial \nu^i}&=& \; \sum_j C^4_{ij} \nu^j+ \sum_j C^5_{ij} L^j_{\alpha} H_\beta \eps ~,\\
\frac{\partial W_{\rm neutrinos}}{\partial L^j_{\alpha}}&=& \; \sum_i C^5_{ij} \nu^i  H_\beta \eps ~.
\eea
Gathering these extra terms with the contributions from the minimal superpotential~\eqref{minimalew} gives the full set of F-term equations:
\bea \label{vernH}
\sum_{i,j} C^5_{ij} \nu^i L^j_{\alpha}\eps-C^0 \barH_\alpha \eps&=& 0\; ,\\ \label{vernHb}
C^0 H_\alpha \eps + \sum_{i,j} C^3_{ij} e^i L^j_{\alpha} \eps&=& 0\; ,\\ \label{vernL}
\sum_i C^5_{ij} \nu^i  H_\beta \eps+\sum_i C^3_{ij} e^i \barH_\beta \eps&=& 0\; ,\\ \label{vernnu}
\sum_j C^4_{ij} \nu^j+ \sum_j C^5_{ij} L^j_{\alpha} H_\beta \eps&=& 0\; ,\\ \label{verne}
\sum_i C^3_{ij} L^j_{\alpha} \barH_\beta \eps&=& 0\; .
\eea
Recently in~\cite{He:2014loa}, it has been shown that this system of equations implies that the following GIOs vanish:
\be
\nu^i=0\; , \quad
LH=0\; , \quad
H\barH=0\; , \quad
L\barH e=0\;.
\ee
Moreover, the only non-trivial equation remaining is~(\ref{vernHb}).
Contracting this condition with $L^k_\beta$, we obtain:
\be\label{con}
\sum_{i,j} C^3_{ij} e^i L^j_{\alpha}L^k_\beta \eps= 0\; .
\ee
The vacuum geometry is therefore given by the relations and syzygies of the $LLe$ operators intersected with the hypersurface defined by~\eqref{con}.

For the sake of completeness, let us recall the demonstration from~\cite{He:2014loa}.
First, from~(\ref{verne}) and from the non-singularity of the coupling matrix $C^3_{ij}$, we conclude that the GIOs $L\barH e$ must all vanish.
Second, we contract~(\ref{vernL}) with $L^k_\alpha$ to obtain:
\be
\sum_i C^5_{ij} \nu^i L^k_{\alpha} H_\beta \eps+ \sum_i C^3_{ij} e^i L^k_{\alpha} \barH_\beta \eps=0\; .
\ee
The second term vanishes by virtue of $L\barH=0$, and, assuming a generic matrix $C^5$, we deduce
\be
\nu^i L^k_{\alpha} H_\beta \eps=0\; .
\ee
This implies that both $\nu^i$ and $LH$ operators vanish.
Indeed, if $\nu^i\neq0$, then $L^k_{\alpha} H_\beta \eps=0$.
From~(\ref{vernnu}), we conclude that $\nu^i=0$, in contradiction with the starting hypothesis.
Therefore $\nu^i=0$, which implies $LH=0$ from~(\ref{vernnu}).
Finally, from~(\ref{vernH}), we also have $\barH=0$. This analysis holds for any number $N_f$ of flavors.

\subsection{Vacuum Geometry}

Let us examine the geometries thus obtained.
We start with the simple case of two flavors as an appetizer.
Then we look at the famous Veronese solution and the corresponding higher-dimensional variety resulting from removing the Majorana mass terms.
Then, we describe the corresponding vacuum geometry for the case of an additional flavor, $N_f=4$.

\subsubsection*{$N_f=2$}

Let us first consider the case of two particle flavors as a warm up. We have seen that it has only two $LLe$ operators coming from the two $e^i$ fields. Moreover, we now have two right-handed neutrino fields.

Considering the case without Majorana mass term, the system of equations~\eqref{con} reduces to:
\bea\label{conexplicit2gen}
C^3_{11}e^1L^1_\alpha L^2_\beta\eps+C^3_{21}e^2L^1_\alpha L^2_\beta\eps=0\, ,\\
C^3_{12}e^2L^1_\alpha L^2_\beta\eps+C^3_{22}e^2L^1_\alpha L^2_\beta\eps=0\, .
\eea
For the case when the coupling matrix $C^3$ is a non-singular matrix, it is clear that the only solution to the above system is when $LLe=0$. Thus the vacuum geometry consists of the point at the origin in $\IC^2$.

Another way to write this ideal is to make the following change of variables:
\be
 \eB_j:= \sum_iC^3_{ij} e^i.
\ee
We subsequently define $y$ variables in the same way as in Section~\ref{s:mutli}:
\be
y_1=\eB_1L^1_\alpha L^2_\beta\eps\, , \quad
y_2=\eB_2L^1_\alpha L^2_\beta\eps\, .
\ee
Thus, the relations~\eqref{conexplicit2gen} immediately become
\be
y_1=0 \, , \quad y_2 =0 \, ,
\ee
and so we indeed have the point at the origin in $\IC^2$ as the vacuum moduli space.

\subsubsection*{$N_f=3$}

For the case of three flavor generations, it is well established that the vacuum geometry is given by the Veronese surface. An analytic demonstration of the Veronese description has recently been obtained in~\cite{He:2014loa}.

To be complete, let us present again the defining polynomial ideal.
Again, it is convenient to make the change of field variables:
\be
 \eB_j:= \sum_iC^3_{ij} e^i ~,
\ee
and define the following $y$ variables:
\be \label{yVeron}
y_{I+C(N_f,2)\cdot (k-1)}=(-1)^{k-1}L^i_\alpha L^j_\beta \eB_k \eps\; .
\ee
With this notation, the $LL\eB$ relations retain a similar form to those we have found in Section~\ref{secminideal}.
In addition, we now have the relation~\eqref{con} which corresponds to,
\bea
y_1-y_9=0 \; ,\\
y_2-y_6=0 \; ,\\
y_4-y_8=0 \; .
\eea
Thus, the full ideal is given by,
\bea\label{Veronese-ideal}\nn
\langle\ y_1y_5-y_2y_4,\; y_1y_6-y_3y_4,\; y_2y_6-y_3y_5,\\  y_1y_8-y_2y_7,\; y_1y_9-y_3y_7,\; y_2y_9-y_3y_8,\\ \nn y_4y_8-y_5y_7,\; y_4y_9-y_6y_7,\; y_5y_9-y_6y_8,\\ \nn y_1-y_9,\;y_2-y_6,\;y_4-y_8\ \rangle \; .
\eea

In fact, the last three linear terms can simply be used as constraints within the 9 quadratics and, thus, reduce the ideal as a set of 6 quadratic polynomials. We have,
\be
\cM = (5|3,4|2^6) \; ,
\ee
and the corresponding Hilbert series,
\be
\frac{1+3t}{(1-t)^3} \; .
\ee
It should be noted that the Hilbert series is not palindromic and therefore the geometry is not Calabi--Yau.

The Veronese surface is an embedding of $\IP^2$ into $\IP^5$. It is in fact the only Severi variety on projective dimension two, and it is remarkable that two of the four Severi varieties appear as vacuum geometry for supersymmetric models with three flavor generations. The embedding is explicitly given by:
\be
\ba{ccc}
\IP^{2} & \rightarrow & \IP^{5} \cr
[x_0:x_1:x_2] & \mapsto &
[{x_0}^2 : x_0x_1 : {x_1}^2 : x_0x_2 : x_1x_2 : {x_2}^2]\label{veronese}
\ea
\ee
This can again be understood in terms of the following change of variables:
\be\label{verovar}
\ba{lll}
y_1 \rightarrow x_0x_2 ~, & y_2 \rightarrow x_0x_1 ~, & y_3 \rightarrow x_0^2 ~, \\
y_4 \rightarrow x_1x_2 ~, & y_5 \rightarrow x_1^2 ~, & y_6 \rightarrow x_1x_0 ~, \\
y_7 \rightarrow x_2^2 ~, & y_8 \rightarrow x_2x_1 ~, & y_9 \rightarrow x_2x_0 ~.
\ea
\ee
It should be observed that the effect of~\eqref{con} is therefore to identify the two projective spaces arising from the Grassmannian $Gr(3,2)$ and $\IP^2$. Imposing the identification relation $[x_0:x_1:x_2] = [z_0:z_1:z_2]$ onto~\eqref{Segre} lead to the vacuum geometry in the presence of right-handed neutrinos.

Again, we see from the binomial nature of the polynomial ideal~\eqref{Veronese-ideal} that the Veronese variety is toric. Using the same notation as previously, the corresponding diagram is given by,
\begin{equation}
\kN =
\left(
\begin{array}{ccc}
 -1 & -2 & 0 \\
 1 & 0 & 0 \\
 0 & -1 & 0 \\
 1 & 0 & -2 \\
 1 & 0 & -1 \\
 0 & -1 & -1 \\
\end{array}
\right) \qquad \Longrightarrow
\qquad
\begin{array}{c}
\includegraphics[width=7cm]{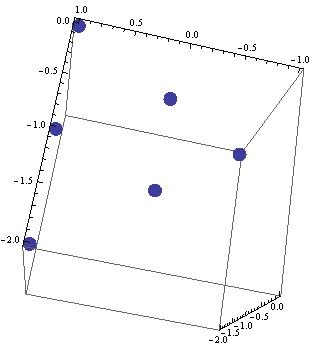}
\end{array}
\end{equation}
where we could include a pictorial representation of the toric cone, as it sits within three dimensions.

For the base space $\CB$ of the affine cone, we can compute its Hodge diamond
\begin{equation}
h^{p,q}(\CB) \quad = \quad
{\begin{array}{ccccc}
&&h^{0,0}&& \\
&h^{0,1}&&h^{0,1}& \\
h^{0,2}&&h^{1,1}&&h^{0,2} \\
&h^{0,1}&&h^{0,1}& \\
&&h^{0,0}&& \\
\end{array}}
\quad = \quad
{\begin{array}{ccccc}
&&1&& \\
&0&&0& \\
0&&1&&0 \\
&0&&0& \\
&&1&& \\
\end{array}}.
\label{hodge1}
\end{equation}
This confirms our identification of the Veronese geometry.

\subsubsection*{$N_f=4$}

We give only the Hilbert series and dimension as writing the full ideal is tedious and ultimately unilluminating. However we realize that all the previous structures we have noted remain the same. The geometry stems out of $Gr(4,2) \times \IP^3$. As in the Veronese case, some identifications occur between the points in $Gr(4,2)$ and $\IP^3$ due to~\eqref{con}. With the $y$ variables definition~\eqref{yVeron}, these linear relations~\eqref{con} become:
\bea
y_{1}-y_{16}+y_{23}=0 \; ,\\
y_{2}-y_{10}+y_{24}=0 \; ,\\
y_{3}-y_{11}+y_{18}=0 \; ,\\
y_{7}-y_{14}+y_{21}=0 \; .
\eea
The identification is therefore not as straightforward as for the Veronese case, since we have a sum of three terms in each equality. In terms of the $Gr(4,2)$ and $\IP^3$ variables, keeping the same variables as defined by~\eqref{newvar4}, these linear equations become
\bea\label{identGr42C4-1}
z_0x_0-z_2x_3+z_3x_4=0 \; ,\\\label{identGr42C4-2}
z_0x_1-z_1x_3+z_3x_5=0 \; ,\\\label{identGr42C4-3}
z_0x_2-z_1x_4+z_2x_5=0 \; ,\\\label{identGr42C4-4}
z_1x_0-z_2x_1+z_3x_2=0 \; .
\eea
The vacuum moduli space is therefore given by~\eqref{newvar4}, subject to the constraints~\eqref{Pluck} and~\eqref{identGr42C4-1}--\eqref{identGr42C4-4}.
It corresponds to a geometry of the type
\beq\label{Gr24}
\cM = (19|6,40|2^{84}) ~.
\eeq
It is irreducible, and its Hilbert series is
\be
\frac{1+14t+21t^2+4t^3}{(1-t)^6} ~.
\ee
Again, we do not have a palindromic Hilbert series, so the geometry fails to be Calabi--Yau.
Computationally, identifying further geometrical invariants such as Euler number or Hodge numbers for this space, due to the complexity of the defining ideal, is prohibitively lengthy on standard desktop computers. Nonetheless, it is not without hope that future advances in algebraic geometry software packages will make such computations more easily accessible. 

\subsection{Role of the Majorana Mass Term}

It should be noted that key to the argument in the previous subsection is the fact that the $\nu^i$ vanish due to equation~\eqref{vernnu}.
Now, when considering a superpotential without Majorana mass terms, {\em e.g.}, with
$$
C^4=0 \; ,
$$
this argument does not apply anymore.
Instead, we have the following system for the F-term equations:
\bea \label{DvernH}
\sum_{i,j} C^5_{ij} \nu^i L^j_{\alpha}\eps-C^0 \barH_\alpha \eps&=& 0\; ,\\ \label{DvernHb}
C^0 H_\alpha \eps + \sum_{i,j} C^3_{ij} e^i L^j_{\alpha} \eps&=& 0\; ,\\ \label{DvernL}
\sum_i C^5_{ij} \nu^i  H_\beta \eps+ \sum_i C^3_{ij} e^i \barH_\beta \eps&=& 0\; ,\\ \label{Dvernnu}
\sum_j C^5_{ij} L^j_{\alpha} H_\beta \eps&=& 0\; ,\\ \label{Dverne}
\sum_i C^3_{ij} L^j_{\alpha} \barH_\beta \eps&=& 0\; .
\eea
From this, we can deduce again that all $L\barH$ must vanish from~\eqref{Dverne}.
The difference is now that we can also deduce that all $LH$ must vanish from~\eqref{Dvernnu}.
Finally, contracting~\eqref{DvernH} with $H_\beta$ and with $LH=0$, we also have that $H\barH=0$.
In summary, we have the following vanishing GIOs,
\be
LH=0\; , \quad
H\barH=0\; , \quad
L\barH e=0\;.
\ee

Again, we have the conditions~\eqref{con} for the $LLe$ operators.
However, we now have an extra condition coming from contracting equation~\eqref{DvernH} with $L^k_{\beta}e^l$.
This leads to the two constraints:
\bea\label{conD1}
\sum_{i,j} C^3_{ij} e^i L^j_{\alpha}L^k_\beta \eps= 0\; , \\\label{conD2}
\sum_{i,j}C^5_{ij} \nu^i L^j_{\alpha}L^k_{\beta}e^l\eps=0\; ,
\eea
and the geometry is given by the $LLe$ and $\nu$ operators satisfying these conditions.

The extra condition~\eqref{conD2} that we now have for the non-vanishing $\nu$ operators are quadratic polynomials.
They are also the only polynomials involving $\nu$.
The $LLe$ operators are subject to the same constraints as the case with Majorana mass terms.
Therefore, we have an embedding of this geometry onto a higher dimensional algebraic variety incorporating $\nu$ degrees of freedom.
This embedding is however non-trivial, as we will see below.

\subsubsection*{$N_f=2$}

Considering the case without Majorana mass terms for the right-handed neutrinos, we see that we now also need to satisfy~\eqref{conD2}.
The first condition will lead to $LLe=0$ as above and the second condition will thus be trivially satisfied.
The right-handed neutrino fields thus remain unconstrained, and the vacuum moduli space is then $\cM=\IC^2$.

\subsubsection*{$N_f=3$}

Let us now consider the case without the Majorana mass term in the superpotential, when $C^4=0$.
As explained previously, the right-handed neutrinos do not vanish anymore.
Similarly to the $e^i$ fields, we can absorb the coupling constant $C^5$ into a field redefinition,
\be\label{newnu}
 \tilde\nu_j:= \sum_iC^5_{ij} \nu^i\, .
\ee
We can also define the additional $y$ variables,
\be
y_{10}=\tilde\nu_1 \, ,\quad y_{11}=\tilde\nu_2 \, ,\quad y_{12}=\tilde\nu_3\, .
\ee
We must now consider an ideal in $\IC^{12}$. The polynomials from~\eqref{Veronese-ideal} remain part of the defining polynomials for the vacuum geometry.
In addition, we now have the condition~\eqref{conD2}.
This gives,
\bea
\nn
y_{11}y_{1}+y_{12}y_{2}=0\, ,\quad
y_{10}y_{1}-y_{12}y_{3}=0\, ,\quad
y_{10}y_{2}+y_{11}y_{3}=0\, ,\\
\nn
y_{11}y_{4}+y_{12}y_{5}=0\, ,\quad
y_{10}y_{4}-y_{12}y_{6}=0\, ,\quad
y_{10}y_{5}+y_{11}y_{6}=0\, ,\\
y_{11}y_{7}+y_{12}y_{8}=0\, ,\quad
y_{10}y_{7}-y_{12}y_{9}=0\, ,\quad
y_{10}y_{8}+y_{11}y_{9}=0\, .
\eea

The full ideal is then,
\bea\nn
\langle\ y_1y_5-y_2y_4,\; y_1y_6-y_3y_4,\; y_2y_6-y_3y_5,\\ \nn  y_1y_8-y_2y_7,\; y_1y_9-y_3y_7,\; y_2y_9-y_3y_8,\\ \nn y_4y_8-y_5y_7,\; y_4y_9-y_6y_7,\; y_5y_9-y_6y_8,\\ \label{CY4}
y_1-y_9,\;y_2-y_6,\;y_4-y_8,\\ \nn
y_{11}y_{1}+y_{12}y_{2}, \;
y_{10}y_{1}-y_{12}y_{3}, \;
y_{10}y_{2}+y_{11}y_{3}, \\ \nn
y_{11}y_{4}+y_{12}y_{5}, \;
y_{10}y_{4}-y_{12}y_{6}, \;
y_{10}y_{5}+y_{11}y_{6}, \\ \nn
y_{11}y_{7}+y_{12}y_{8}, \;
y_{10}y_{7}-y_{12}y_{9}, \;
y_{10}y_{8}+y_{11}y_{9} \
 \rangle \; .
\eea
This ideal corresponds to
\beq
\cM = (8|4,7|2^{14})\, .
\eeq
We can see that the last nine polynomials from~\eqref{CY4} are the only ones containing the neutrino field variables $y_{10}$, $y_{11}$ and $y_{12}$.
They also have the property that they all vanish when $y_{10}=y_{11}=y_{12}=0$, thus recovering the Veronese ideal for the particular point in the vacuum where the neutrino fields vanish.
This implies that the Majorana mass terms would simply lift the right-handed neutrinos from the vacuum.

In analogy with the Veronese analysis, we can define the following change of variables:
\be\label{verovar2}
\ba{lll}
y_1 \rightarrow x_0x_2 ~, & y_2 \rightarrow x_0x_1 ~, & y_3 \rightarrow x_0^2 ~, \\
y_4 \rightarrow x_1x_2 ~, & y_5 \rightarrow x_1^2 ~, & y_6 \rightarrow x_1x_0 ~, \\
y_7 \rightarrow x_2^2 ~, & y_8 \rightarrow x_2x_1 ~, & y_9 \rightarrow x_2x_0 ~, \\
y_{10} \rightarrow x_0 \lambda~, & y_{11} \rightarrow x_1 \lambda~, & y_{12} \rightarrow x_2\lambda ~,
\ea
\ee
where we introduced a new coordinate $\lambda$. This change of variables satisfies automatically all constraints form the ideal~\eqref{CY4}. It can be understood as the following embedding,
\begin{equation} \label{Segre2}
\begin{array}{ccccc}
\IP^2 & \times & \IC & \longrightarrow & \IP^8 \\
\mbox{$[x_0 : x_1 : x_2]$}
& &  
\mbox{$[\lambda]$}
& \rightarrow & [{x_0}^2 : x_0x_1 : {x_1}^2 : x_0x_2 : x_1x_2 : {x_2}^2 : x_0\lambda : x_1\lambda : x_2\lambda] \\
\end{array}
\end{equation}
A general treatment of the corresponding embedding for more general cases with $N_f \geq 4$ will be presented in Subsection~\ref{sec-gen}.

Using algebraic geometry packages~\cite{mac,sing}, we can compute its Hilbert series and obtain
\be
\frac{1+5t+t^2}{(1-t)^4} \, .
\ee
Thus the geometry is an (irreducible) non-compact affine Calabi--Yau.
Moreover, we see that the ideal~\eqref{CY4} contains only binomials, thus is toric again.
The removal of the Majorana mass term for the right-handed neutrinos thus brings back this property.
The toric diagram is given by,
\begin{equation}
\kN = \left(
\begin{array}{cccc}
 -2 & 0 & 0 & -1 \\
 0 & 0 & 0 & 1 \\
 -1 & 0 & 0 & 0 \\
 0 & -2 & 2 & 1 \\
 0 & -1 & 1 & 1 \\
 -1 & -1 & 1 & 0 \\
 -1 & 1 & 0 & 0 \\
 0 & 1 & 0 & 1 \\
 0 & 0 & 1 & 1 \\
\end{array}
\right)
\qquad \Longrightarrow \qquad
\begin{array}{c}
\includegraphics[width=7cm]{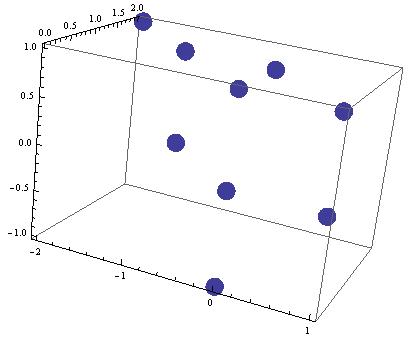}
\end{array}
\end{equation}
where the pictorial representation corresponds to the three-dimensional hyperplane in which all the vectors fit, due to the Calabi--Yau property.

The Hodge diamond of the compact base manifold $\CB$ of the projective variety can be computed.
We find
\begin{equation}
h^{p,q}(\CB) \; = \;
{\begin{array}{ccccccc}
&&&h^{0,0}&&& \\
&&h^{0,1}&&h^{0,1}&& \\
&h^{0,2}&&h^{1,1}&&h^{0,2}& \\
h^{0,3}&&h^{1,2}&&h^{1,2}&&h^{0,3} \\
&h^{0,2}&&h^{1,1}&&h^{0,2}& \\
&&h^{0,1}&&h^{0,1}&& \\
&&&h^{0,0}&&& \\
\end{array}}
\; = \;
{\begin{array}{ccccccc}
&&& 1 &&& \\
&& 0 && 0 && \\
& 0 && 2 && 0 & \\
0 && 0 && 0 && 0  \\
& 0 && 2 && 0 & \\
&& 0 && 0 && \\
&&& 1 &&& \\
\end{array}}\, .
\label{hodge3}
\end{equation}
Similarly as for the five-dimensional vacuum of the minimal superpotential and the Veronese surface, this Hodge diamond has the property to be non-vanishing in its diagonal only. It is consistent with the Hodge diamond of $\IP^2 \times \IP^1$ as can be seen by using the K\"unneth formula. Thus, as with \eqref{hodge2}, our vacuum moduli space is topologically $\IP^2 \times \IP^1$ but algebro-geometrically we can pin-point it as the toric variety given above.

\subsubsection*{$N_f=4$}

Similarly to the $N_f=3$ case, we expect the geometry to be some fibration over the geometry described previously by \eqref{Gr24} in the $N_f=4$ case with the Majorana mass term. Let us introduce the neutrino variables
\be
y_{25}=\tilde\nu_1 \, ,\quad y_{26}=\tilde\nu_2 \, ,\quad y_{27}=\tilde\nu_3\, ,\quad y_{28}=\tilde\nu_4\, .
\ee
where the coupling constant $C^5$ is absorbed into $\tilde\nu$ as in~\eqref{newnu}. Since the Grassmannian $Gr(4,2)$ does not correspond to a projective space, it is not straightforward to give the embedding of the vacuum moduli space. However, we can give the constraint equations for the neutrinos variables which determine the fibration of these extra variables over the geometry described in~\eqref{Gr24}.
From~\eqref{conD2}, we have,
\bea
-y_{26}y_{1+6n}-y_{27}y_{2+6n}-y_{28}y_{3+6n}=0 \; ,\\
y_{25}y_{1+6n}-y_{27}y_{4+6n}-y_{28}y_{5+6n}=0 \; ,\\
y_{25}y_{2+6n}+y_{26}y_{4+6n}-y_{28}y_{6+6n}=0 \; ,\\
y_{25}y_{3+6n}+y_{26}y_{5+6n}+y_{27}y_{6+6n}=0 \; ,
\eea
for $n=0,1,2,3$. These $16$ equations clearly vanish when the neutrino variables are set to zero and we recover the ideal from~\eqref{Gr24}. Thus the effect of the Majorana mass term is simply to lift the neutrinos variables from the vacuum.

The above geometry, however, is not as trivial as for the Veronese case. Here, we have a geometry of the type
\beq
\cM = (23|8,71|3^{11}2^{99}) ~,
\eeq
which is irreducible, and with its Hilbert series given by,
\be
\frac{1+16t+37t^2+16t^3+t^4}{(1-t)^8} ~.
\ee
As the numerator is palindromic, we conclude that the vacuum manifold is Calabi--Yau.

\subsection{Vacuum Geometry: General $N_f$}\label{sec-gen}
Having gained experience with the cases of $N_f=2,3,4$ using algorithmic geometry, we can now analytically study the general case.
We assume that the matrices $C^5_{ij}$ and $C^3_{ij}$ have full rank, and, without loss of generality, that $C^0$ is nonzero. As previously, we set
\bea\label{what}
\tilde{e}_j :=  \sum_{i} C^3_{ij} e^i \; , \\\label{what2}
\tilde{\nu}_j :=  \sum_{i}C^5_{ij} \nu^i \; .
\eea
Now, define three matrices of variables by:
\begin{equation}
{ %
E = 
\left(
\begin{array}{c}
 \tilde{e} \\ \tilde{\nu}  \\
\end{array}
\right) = \frac{1}{C_0}
\left(
\begin{array}{cccc}
 -\tilde{e}_1 & -\tilde{e}_2 & \ldots & -\tilde{e}_{N_f} \\
 \tilde{\nu}_1 & \tilde{\nu}_2 & \ldots & \tilde{\nu}_{N_f} \\
\end{array}
\right)
},\quad
L = \left(
\begin{array}{cc}
 L_1 & L_2  \\
\end{array}
\right) = 
\left(
\begin{array}{cc}
 L_1^1 & L_2^1  \\
 \vdots & \vdots \\
 L_1^{N_f} & L_2^{N_f} \\
\end{array}
\right),\quad
H = \left(
\begin{array}{cc}
 H_1 & H_2  \\
 \barH_1 & \barH_2  \\
\end{array}
\right)\ .
\end{equation}
We will think of the row vectors $\tilde{e}$ and $\tilde{\nu}$ as hyperplanes, and the column vectors $L_1$ and $L_2$ as points in an affine or projective space.

With this notation, the four equations~\eqref{DvernH} and \eqref{DvernHb} are equivalent to the matrix equation 
\begin{equation}
H = E L ~;
\end{equation}
the $2N_f$ equations~\eqref{DvernL} translate to the matrix equation
\begin{equation}
{ %
H^T \left(
\begin{array}{cc}
 0 & 1 \\
 -1 & 0 \\
\end{array}
\right)
} E = 0 ~;
\end{equation}
and the $2N_f$ equations~\eqref{Dvernnu} and~\eqref{Dverne} translate to
\begin{equation}
{ %
H \left(
\begin{array}{cc}
 0 & 1 \\
 -1 & 0 \\
\end{array}
\right)
} L^T = 0 ~.
\end{equation}
Eliminating the variables in $H$ by using the first of these equations leaves $4N_f$ equations in the variables $\tilde{e}_j$, $\tilde{\nu}_j$, and $L^j_\alpha$:
\begin{equation}\label{LEE}
{ %
L^T E^T \left(
\begin{array}{cc}
 0 & 1 \\
 -1 & 0 \\
\end{array}
\right)
} E = 0 \ ,
\qquad
{ %
E L \left(
\begin{array}{cc}
 0 & 1 \\
 -1 & 0 \\
\end{array}
\right)
} L^T = 0 .
\end{equation}
These equations are homogeneous separately in the four sets of $N_f$
variables: $\tilde{e}_j$, $\tilde{\nu}_j$, $L^j_1$, and $L^j_2$, so
let $X \subset \IP^{N_f-1} \times \IP^{N_f-1} \times \IP^{N_f-1}
\times \IP^{N_f-1}$ be the zero set of these $4N_f$ equations, with each
factor of projective space parametrized by one of the four sets of
variables.

As we are interested in the vacuum moduli space, which is the affine cone over the image of $X$
under the GIO's $\nu$ and $LLe$, let $\Delta_{ij} = L^i_\alpha L^j_\beta \eps$ be the $2 \times 2$
minors of the matrix $L$, and let $\Delta$ be the ideal generated by these 
minors.  Any point on $V(\Delta)$, where $V$ denotes the variety corresponding to the ideal,  maps via the $LLe$ operators to the origin, so the images of points in $X \cap V(\Delta)$
are easy to understand, and are subvarieties of the moduli spaces identified below.

Let us compute equations which cut out $X \setminus V(\Delta)$.  
Multiply the second matrix equation~\eqref{LEE} by the $N_f \times 2$ matrix which is zero except in
rows $i$ and $j$: the $i$-th row is $(-L_j^1, -L_j^2)$, and the $j$-th row is $(L_i^1, L_i^2)$, obtaining
\begin{equation}
{ %
0 = E L \left(
\begin{array}{cc}
 0 & 1 \\
 -1 & 0 \\
\end{array}
\right)
} L^T
\left(
\begin{array}{cc}
 0 & 0 \\
 \vdots & \vdots \\
 -L_j^1 & -L_j^2 \\
 \vdots & \vdots \\
 L_i^1 & L_i^2 \\
 \vdots & \vdots \\
 0 & 0 \\
\end{array}
\right) = \Delta_{ij} E L 
\end{equation}
Since this holds for all $i$ and $j$, dividing by $\Delta_{ij}$ we see that $X \setminus V(\Delta)$ is cut out by the four polynomial entries of the
matrix $EL$.  Notice that once these vanish, then certainly the matrix equations~\eqref{LEE} vanish.
On the complement of $V(\Delta)$, these four polynomials define an irreducible variety of codimension four.
(If, for instance, we consider the subset for which $\Delta_{12} \neq 0$, we can multiply the matrix $L$ on the right to ensure that the first 2 rows of $L$
form the identity matrix.  The resulting four equations write $\tilde{e}_1, \tilde{e}_2, \tilde{\nu}_1, \tilde{\nu}_2$ in terms of the other variables,
which shows that the variety is irreducible of codimension four.)
Therefore, the closure
$Y$ of $X \setminus V(\Delta)$ is also irreducible with codimension four (this holds for any $N_f \geq 2$, although it is not very interesting
for $N_f = 2$).  Therefore, $\dim Y = 4(N_f-1) - 4 = 4N_f - 8$.  A closer analysis using {\it Macaulay2}~\cite{mac}
shows that for $N_f \geq 4$, the ideal of $Y$ is generated by these four polynomials.  For $N_f = 3$,
the ideal is generated by these four polynomials, together with the $2 \times 2$ minors of the matrix $E$.

Notice that in the case $N_f \geq 4$, $Y$ can be described more geometrically as the locus of 
\begin{equation}
(\tilde{e}, \tilde{\nu}, L_1, L_2) \in \IP^{N_f-1} \times \IP^{N_f-1} \times \IP^{N_f-1} \times \IP^{N_f-1}
\end{equation}
such that the hyperplanes $\tilde{e}$ and $\tilde{\nu}$ contain the points $L_1$ and $L_2$ (and therefore the line $M$ joining $L_1$ and $L_2$, if these points are distinct).

For $N_f = 3$, $Y$ is described geometrically as the locus of
\begin{equation}
(\tilde{e}, \tilde{\nu}, L_1, L_2) \in \IP^{2} \times \IP^{2} \times
\IP^{2} \times \IP^{2}
\end{equation}
such that the {\it lines} $\tilde{e}$ and
$\tilde{\nu}$ are equal, and contain the points $L_1$ and $L_2$ (and
therefore equals the line $M$ joining $L_1$ and $L_2$, if these points are
distinct).

Now, consider the image of $Y$ under the GIO's $LLe$ and $\nu$.  
This map factors as follows:
\begin{equation}\label{Y12}
\begin{array}{ccccc}
\IP^{N_f-1}_\nu \times \IP^{N_f-1}_e \times \IP^{N_f-1}_{L_1} \times \IP^{N_f-1}_{L_2} & \longrightarrow & \IP^{N_f-1}_\nu \times \IP^{N_f-1}_e \times Gr(2,N_f) 
 & \longrightarrow & \IP^{N_f-1}_\nu \times \IP^{{N_f \choose 2} N_f - 1} 
\\
\cup & & \cup & & \cup \\
Y & \longrightarrow & Y_1 & \longrightarrow & Y_2 
\end{array}
\ ,
\end{equation}
where the first map is given by the minors of the matrix $L$, and is the identity on the first two factors.  The second map is the Segr\`e embedding.
As marked, let $Y_1$ be the image of $Y$ under the first map, and let $Y_2$ be the image under the final map.  The fibers of the map $Y \longrightarrow Y_1$ 
have dimension two, and therefore the dimension of $Y_1$ is $4N_f-10$.  The second map is an isomorphism of $Y_1$ and $Y_2$, and so
they have the same dimension.  
In conclusion, for $N_f \ge 3$, the vacuum moduli space $\cM$ is the affine cone over $Y_2$, and so has dimension two larger, giving in general that
\begin{equation}
\dim \cM = 4N_f-8 \ , \qquad N_f \ge 3 \ .
\end{equation}

The locus $Y_1$, in the case $N_f \geq 4$, is described geometrically as the set of
\begin{equation}
(\tilde{e}, \tilde{\nu}, M) \in \IP^{N_f-1}_e \times \IP^{N_f-1}_\nu \times Gr(2, N_f)
\end{equation}
such that the hyperplanes $\tilde{e}$ and $\tilde{\nu}$ contain the line $M$.
In the case $N_f = 3$, $Y_1$ is the set
\begin{equation}
(\tilde{e}, \tilde{\nu}, M) \in \IP^{2} \times \IP^{2} \times Gr(2,3)
\end{equation}
such that the lines $\tilde{e}$,
$\tilde{\nu}$, and $M$ are all equal.

\section{Discussion and Outlook}\label{conclusion}\setall

In order to fully appreciate the vacuum moduli space geometries' dependence on the electroweak theories --- that is, the field content and superpotential --- let us tabulate a summary of all the results previously described.
On physical grounds, we know that three generations of Standard Model matter fields are required for CP violation.
On geometric grounds, the vacuum of the electroweak sector is trivial when $N_f < 3$.
Thus, in the table below, we omit the cases of $N_f=2$, which give points or $\IC^2$.
The table lists the GIOs that are non-vanishing in the vacuum.
The toric property refers to whether the ideals are explicitly in a toric form.
The Calabi--Yau property is checked by the palindromicity of the numerator of the Hilbert series associated to the geometry $\cM$.
\begin{table}[h!]
{\begin{center}\begin{footnotesize}
\begin{tabular}{|c|c|c||c|c|c|c|}\hline
\mbox{W} & \mbox{Vacuum GIOs} & $N_f$  & \mbox{dimension} & \mbox{degree} & \mbox{Toric} & \mbox{Calabi--Yau}\\
\hline \hline
$H\barH+L\barH e$ & $LLe, \ LH$  & 3 $\star$ &  5 & 6 & \checkmark & \checkmark \\\cline{3-7}
  & & 4 &  8 & 70 &  & \checkmark \\\cline{3-7}
  & & 5 &  11 & 1050 &  & \checkmark \\\hline
$H\barH+L\barH e+LH\nu+\nu\nu$ & $LLe$ & 3 $\dagger$ & 3 & 4 & \checkmark &  \\\cline{3-7}
  && 4 &   6 & 40 &  &  \\\hline
$H\barH+L\barH e+LH\nu$ & $LLe, \ \nu$ & 3   & 4 & 7 & \checkmark & \checkmark \\\cline{3-7}
  && 4 & 8 & 71 &  & \checkmark \\\hline
\end{tabular}
\end{footnotesize}\end{center}}{\caption{\label{tbl:gio}
{\sf Summary of algebraic geometries encountered as the vacuum moduli space of supersymmetric electroweak theories. Here $W$ is the superpotential; vacuum GIOs are the GIOs after imposing the F-terms, and thus furnish explicit coordinates of the moduli space, of affine dimension and degree as indicated; $N_f$ is the number of generations. We also mark with ``\checkmark'' if the vacuum moduli space is toric or Calabi--Yau.
Furthermore, the $\dagger$ corresponds to the cone over the Veronese surface and the $\star$, the Segr\`e variety. These two are Severi varieties, in fact, the only two which are isomorphic to (products of) projective spaces.
}}}
\end{table}

The observations that can be drawn from this table are the following.
First, for the minimal superpotential, that is $W=H\barH+L\barH e$, the dimension increases by three when adding one more flavor in the theory (which corresponds to adding three fields). The geometries for this superpotential correspond to the affine cone over $Gr(N_f,2)\times \IP^{N_f-1}$, which is a Calabi--Yau space.
Therefore, the affine dimension \footnote{
Incidentally, we note that the degree is $\frac{(3N_f)!}{N_f!^3 (9N_f-3)}$, which happens to be~\cite{oeis} the number of possible necklaces consisting of $N_f$ white beads, $N_f$ red beads and $N_f-1$ black beads, where two necklaces are considered equivalent if they differ by a cyclic permutation. This is due to the Grassmannian symmetry.
}
of the moduli space is $3N_f - 4$.

With the addition of the right handed neutrino in the superpotential, i.e., 
$W=H\barH+L\barH e + LH\nu$, the geometry becomes an affine cone over the bi-projective variety $Y_2$ described in \eqref{Y12}, of affine dimension $4N_f - 8$.
Alternatively, the moduli space can be seen as a double affine cone over a complete intersection in $(\IP^{N_f-1})^4$. For $N_f=3$, it is topologically a cone over $\IP^2 \times \IP^1$.
In any event, the moduli space is Calabi--Yau.
However, when we further we add the Majorana mass term for the neutrino, giving the superpotential $W=H\barH+L\barH e + LH\nu + \nu\nu$, this lifts the neutrinos variables form the vacuum, having the effect of removing the Calabi--Yau property of the vacuum. In particular, for $N_f=3$, we have the cone over the Veronese surface.

One intriguing observation can be readily made: we have shown that {\it only for three generations} do we obtain toric varieties for all superpotentials considered. 
Moreover, at $N_f=3$, we obtain two of the four Severi varieties as the vacuum moduli space:
the cone over the Veronese for $W=H\barH+L\barH e + LH\nu + \nu\nu$ and the cone over the Segr\`e variety for $W=H\barH+L\barH e$. These unique Severi varieties of dimension two and four are, in fact, the only Severi varieties which are themselves projective.
It is interesting that this ``triadophilia'' --- the love of three generations of particles --- could so be geometrically interpreted; one could compare and contrast with~\cite{Candelas:2007ac} for the context of this ``threeness'' in string compactification.

An increase in the number of flavors introduces non-binomial constraints in the variety ideals. It would be worth investigating these varieties from an algebraic geometry point of view to understand whether any properties relate them together, such as in the case of the Severi varieties.
We should also be mindful of~\eqref{t:severi}, especially of the underlying Lie group structure of these spaces. Because we have obtained the first two Severi varieties which are essentially complex projective spaces and which have $S(U(1) \times U(2))$ isometry, it is conceivable that they arise because of the electroweak gauge group. It will be indeed interesting to see whether the other two arise for other gauge groups.
These await further computations.

A full categorization of the vacuum moduli spaces obtained with all possible combinations of the renormalizable terms in the superpotential is under way.
The algorithmic complexity of Gr\"obner bases decomposition render some computations out of reach of personal computers.
However, it is not without hope that a numerical approach might lead to a complete calculation of all the different possibilities for the electroweak sector of supersymmetric theories.

\section*{Acknowledgements}
We thank Noah Daleo, James Gray, Jonathan Hauenstein, and Dhagash Mehta for past, present, and future collaborations on similar themes.
YHH is indebted to the Science and Technology Facilities Council, UK, for grant ST/J00037X/1, the Chinese Ministry of Education, for a Chang-Jiang Chair Professorship at NanKai University, and the city of Tian-Jin for a Qian-Ren Award.
He also thanks Sheldon Katz and Hal Schenck for many fun discussions.
VJ is supported by the South African Research Chairs Initiative of the Department of Science and Technology and National Research Foundation and thanks KIAS and the Yukawa Institute for hospitality during the concluding stages of this project.
CM is grateful to the Helios Foundation for financial support, to Damien Matti for providing computing resources, and Hwasung Lee for useful discussions.
{\it Ioanne, Thomas, Maria et Catherina, filiis dilectis, atque Sanctis patronis eorundem
nominum, et ad maiorem Dei gloriam, Brentus Nelson et Evaristus Y-H He hoc opusculum
dedicant.}
The work of YHH, VJ, and BDN is partially funded by the U.S.\ National Science Foundation under the grant CCF-1048082, EAGER: CiC: A String Cartography. The work of MS is partially funded by the U.S. National Science Foundation under the grant DMS-1002210.

~\\
~\\

\renewcommand{\baselinestretch}{1}

\newpage

\appendix

\section{Gauge Invariant Operators in the MSSM}\label{GIOs}

\begin{table}[h!!!]
{\begin{center}
\begin{tabular}{|c||c|c|c|}\hline
\mbox{Type} & \mbox{Explicit Sum} & \mbox{Index} & \mbox{Number} \\
\hline \hline
$LH$  & $L^i_\alpha H_\beta \eps$ & $i=1,2,3$ & 3 \\ \hline
$H\barH$ & $H_\alpha \barH_\beta \eps$ & & 1  \\ \hline
$udd$ & $u^i_a d^j_b d^k_c \epsilon^{abc}$ & $i,j=1,2,3$; $k=1,\ldots,j-1$ & 9  \\ \hline
$LLe$ & $L^i_\alpha L^j_\beta e^k \eps$ & $i,k=1,2,3$; $j=1,\ldots,j-1$ & 9  \\ \hline
$QdL$ & $Q^i_{a, \alpha} d^j_a L^k_\beta \eps$ & $i,j,k=1,2,3$ & 27 \\ \hline
$QuH$ & $Q^i_{a, \alpha} u^j_a H_\beta \eps$ & $i,j=1,2,3$ & 9 \\ \hline
$Qd\barH$ & $Q^i_{a, \alpha} d^j_a \barH_\beta \eps$ & $i,j=1,2,3$ & 9 \\ \hline
$L\barH e$ & $L^i_\alpha \barH_\beta \eps e^j$ & $i,j=1,2,3$ & 9 \\ \hline
$QQQL$ & $Q^i_{a, \beta} Q^j_{b, \gamma} Q^k_{c, \alpha} L^l_\delta \epsilon^{abc} \epsilon^{\beta\gamma}\epsilon^{\alpha\delta}$ & $\ba{l} i,j,k,l=1,2,3; i\ne k, j\ne k, \\ j \le i, (i,j,k) \ne (3,2,1) \ea$ & 24 \\ \hline
$QuQd$ & $Q^i_{a, \alpha} u^j_a Q^k_{b, \beta} d^l_b \eps$ & $i,j,k,l=1,2,3$ & 81 \\ \hline
$QuLe$ & $Q^i_{a, \alpha} u^j_a L^k_{\beta} e^l \eps$ & $i,j,k,l=1,2,3$ & 81 \\ \hline
$uude$ & $u^i_a u^j_b d^k_c e^l \epsilon^{abc}$ & $i,j,k,l=1,2,3; j<i$ & 27 \\ \hline
$QQQ\barH$ & $Q^i_{a, \beta} Q^j_{b, \gamma} Q^k_{c, \alpha} \barH_\delta \epsilon^{abc} \epsilon^{\beta\gamma} \epsilon^{\alpha\delta}$ & $\ba{l} i,j,k=1,2,3; i\ne k, j\ne k, \\ j\le i, (i,j,k) \ne (3,2,1) \ea$ & 8 \\ \hline
$Qu\barH e$ & $Q^i_{a, \alpha} u^j_a \barH_\beta e^k \eps$ & $i,j,k =1,2,3$ & 27 \\ \hline
$dddLL$ & $d^i_a d^j_b d^k_c L^m_\alpha L^n_\beta \epsilon^{abc} \epsilon_{ijk} \eps$ & $m,n=1,2,3, n<m$ & 3 \\ \hline
$uuuee$ & $u^i_a u^j_b u^k_c e^m e^n \epsilon^{abc} \epsilon_{ijk}$ & $m,n=1,2,3, n \le m$ & 6 \\ \hline
$QuQue$ & $Q^i_{a, \alpha} u^j_a Q^k_{b, \beta} u^m_b e^n \eps$ & $ \ba{l} i,j,k,m,n=1,2,3; \\ \mbox{antisymmetric}\{ (i,j), (k,m) \}
\ea$ & 108  \\ \hline
$QQQQu$ & $Q^i_{a, \beta} Q^j_{b, \gamma} Q^k_{c, \alpha} Q^m_{f,\delta} u^n_f \epsilon^{abc} \epsilon^{\beta\gamma} \epsilon^{\alpha\delta}$ & $\ba{l} i,j,k,m,n=1,2,3; i\ne k, j\ne k, \\ j\le i, (i,j,k) \ne (3,2,1) \ea$ & 72  \\ \hline
$dddL\barH$ & $d^i_a d^j_b d^k_c L^m_\alpha \barH_{\beta} \epsilon^{abc}\epsilon_{ijk} \eps$ & $m=1,2,3$ & 3 \\ \hline
$uudQdH$ & $u^i_a u^j_b d^k_c Q^m_{f, \alpha}d^n_f H_\beta \epsilon^{abc} \eps$ & $i,j,k,m,n=1,2,3; j<i$ & 81 \\ \hline
$(QQQ)_4LLH$  & $(QQQ)_4^{\alpha\beta\gamma} L^m_\alpha L^n_\beta H_\gamma$ & $m,n=1,2,3, n \le m$ & 6 \\ \hline
$(QQQ)_4LH\barH$ & $(QQQ)_4^{\alpha\beta\gamma} L^m_\alpha H_\beta \barH_\gamma$ & $m=1,2,3$ & 3 \\ \hline
$(QQQ)_4H\barH\barH$ & $(QQQ)_4^{\alpha\beta\gamma} H_\alpha \barH_\beta \barH_\gamma$ & & 1 \\ \hline
$(QQQ)_4LLLe$ & $(QQQ)_4^{\alpha\beta\gamma} L^m_\alpha L^n_\beta L^p_\gamma e^q$ & $m,n,p,q=1,2,3, n\le m,p \le n$ & 30 \\ \hline %
$uudQdQd$ & $u^i_a u^j_b d^k_c Q^m_{f, \alpha}d^n_f Q^p_{g, \beta} d^q_g \epsilon^{abc} \eps$ & $\ba{l} i,j,k,m,n,p,q=1,2,3; \\ j<i, \mbox{antisymmetric}\{(m,n), (p,q)\} \ea$ & 324 \\ \hline
$(QQQ)_4LL\barH e$ & $(QQQ)_4^{\alpha\beta\gamma} L^m_\alpha L^n_\beta \barH_\gamma e^p$ & $m,n,p = 1,2,3, n \le m$ & 18 \\ \hline %
$(QQQ)_4L\barH\barH e$ & $(QQQ)_4^{\alpha\beta\gamma} L^m_\alpha \barH_\beta \barH_\gamma e^n$ & $m,n=1,2,3$ &  9 \\ \hline
$(QQQ)_4\barH\barH\barH e$ & $(QQQ)_4^{\alpha\beta\gamma} \barH_\alpha \barH_\beta \barH_\gamma e^m$ & $m=1,2,3$ &  3 \\ \hline
\end{tabular}
\end{center}}{\caption{\label{tbl:giofull}
{\sf The set $D = \{r_i\}$ of generators of gauge invariant operators for the  MSSM}.}}
\end{table}

\section{Toric Varieties}\label{toric}

In this appendix, we would like to present how the toric diagrams have been obtained for the relevant varieties.
The reader may find this brief review of use since it differs from most of the toric literature in the physics community, because it emphasizes the Gr\"obner basis and algorithmic aspects of the geometry.  Further clarification of what it exactly means to be Calabi--Yau will be addressed in the next appendix.

The encountered ideals had the property to be a {\bf binomial ideal}, \textit{i.e.}, it consists only of generators of the form of ``monomial'' = ``monomial'',
\begin{equation}
\vec{y}^{\vec{m}_+} = \vec{y}^{\vec{m}_-} \ ,
\quad
\vec{m}_+, \ - \vec{m}_- \in \IZ^k_{\ge 0} \ , y_{j=1,\ldots,k} \in \IC \ ,
\quad
\end{equation}
where the notation $\vec{y}^{\vec{m}}$ denotes the monomials $y_1^{m_1} y_2^{m_2} \ldots y_k^{m_k}$.
Now, an irreducible binomial ideal geometrically describes a toric variety~\cite{toricbook}.
This fact is exploited and constitutes the bipartite structure of toric quiver gauge theories~\cite{He:2012js}.

For the case of the Veronese, the given ideal definition~\eqref{Veronese-ideal} clearly has three redundant {\it linear} variables which simply amount to setting one coordinate variable equal to another, \textit{e.g.}, $y_1 = y_9$.
We strip our ideal of these, giving us what is known as a {\bf minimal presentation} of the ideal as nine quadratics in $\{y_3, y_5, y_6, y_7, y_8, y_9 \}$, which can be seen as projective coordinates of $\IP^5$.
Furthermore, these nine quadratics are not independent and can be generated by only six {\bf minimal generators} in the form of six quadratics which we will see in~\eqref{veroneseToric}.

It is expedient to draw the toric diagram since, after all, within the pictorial lies the power of toric geometry.
The diagram is readily constructed from the exponent vectors in the binomial ideal.
This is done as follows.
First, we extract the exponent vectors. With relative minus signs, we can choose the left-hand side to be $m_+$ and the right-hand side to be $m_-$.
Using the minimal presentation of the binomial ideal, which we now rewrite for the readers' convenience, we obtain:
\begin{equation}\label{veroneseToric}
\begin{array}{ccc}
\begin{array}{c}
\langle
y_{6}\ y_{8}-y_{5}\ y_{9},\, y_{3}\ y_{8}-y_{6}\ y_{9},\,y_{6}\ y_{7}-y_{8}\ y_{9},
\\
y_{5}\ y_{7}-y_{8}^2,\, y_{3}\ y_{7}-y_{9}^2,\, y_{3}\ y_{5}-y_{6}^2
\rangle
\end{array}
&
\Rightarrow
&
\kM :=
\left(\begin{array}{cccccc}
y_3 & y_5& y_6& y_7& y_8& y_9 \\ \hline
0 & -1 & 1 & 0 & 1 & -1 \\
1 & 0 & -1 & 0 & 1 & -1 \\
0 & 0 & 1 & 1 & -1 & -1 \\
0 & 1 & 0 & 1 & -2 & 0 \\
1 & 0 & 0 & 1 & 0 & -2 \\
1 & 1 & -2 & 0 & 0 & 0 \\
\end{array}\right)
\end{array}
\end{equation}
where each row of $\kM$ corresponds to a generator in the (minimally generated and minimally presented) ideal.

Next, we find the relations among these generators, \textit{i.e.}, whether rows obey sum relations.
In other words, we find the {\bf integer kernel} of $\kM$,
which is the matrix of {\it lattice generators} $\kN$ such that $\kM \cdot \kN = 0$.
This is not just the null space of $\kM$ over $\IZ$ but the minimal generators over $\IZ$ of the nullspace, again, an algorithm conveniently implemented in~\cite{mac}.
As a familiar example, consider the {\it conifold}, given by the quadric $uv=zw$ in $\IC^4$.
Here, $\kM = (1,1,-1,-1)$, so the integer kernel is
$\kN= ${\tiny $
\left(
\begin{array}{ccc}
 1 & 1 & 0 \\
 -1 & 0 & 1 \\
 0 & 1 & 0 \\
 0 & 0 & 1 \\
\end{array}
\right)$},
which is the familiar co-planar square cone of the toric diagram for the conifold, with rows as three-vectors.

Returning to the case of the Veronese, we find that
\begin{equation}\label{Nvero}
\kN = \ker_{\IZ}(\kM^T) =
\left(
\begin{array}{ccc}
 -1 & -2 & 0 \\
 1 & 0 & 0 \\
 0 & -1 & 0 \\
 1 & 0 & -2 \\
 1 & 0 & -1 \\
 0 & -1 & -1 \\
\end{array}
\right) \qquad \Longrightarrow
\qquad
\begin{array}{c}
\includegraphics[width=7cm]{veroneseToric.jpg}
\end{array}
\end{equation}
The rows of $\kN$ are automatically of length three, meaning that we can draw them in $\IR^3$, as is indeed required for a three (complex) dimensional toric variety, here, the affine cone over the Veronese surface.
The endpoints are vectors generating the toric cone and we represent them above as lattice points.
The astute reader may question that this looks like the toric diagram for the affine Calabi--Yau singularity $\IC^3 / \IZ_2 \times \IZ_2$.
However this is {\it not} the case: we see that the points in the toric diagram are co-planar at height 2 and not the required height 1 for Calabi--Yau (the reader is referred to Appendix~\ref{ap:toricCY} for a detailed discussion on this point). This is indeed consistent with the fact that the Hilbert series does {\it not} have the palindromic numerator needed for the Gorenstein/Calabi--Yau property.

Using the same methods as above, we can readily obtain the toric diagrams, where possible, of other of our geometries as well.
Of course, if the geometries are of high dimension, say complex dimension $n > 3$, then the toric diagram will consist of lattice points in $\IR^n$ and visualization will become difficult.
Nevertheless let us present them here.

The variety in~\eqref{CY4}, after reducing to minimal presentation and redefining $y_{11} \rightarrow -y_{11}$, indeed corresponds to a binomial ideal of fourteen quadrics in\\ $\IP^8[y_{3}: y_{5}: y_{6}: y_{7}: y_{8}: y_{9}: y_{10}: y_{11}: y_{12}]$, \textit{viz.},
\begin{align}
\nn
\langle &
y_{9}y_{11}-y_{6}y_{12},\, y_{8}y_{11}-y_{5}y_{12},\, y_{7}y_{11}-y_{8}y_{12},\,
y_{9}y_{10}-y_{3}y_{12},\, y_{8}y_{10}-y_{6}y_{12},\, \\
\nn
&y_{7}y_{10}-y_{9}y_{12},\, y_{6}y_{10}-y_{3}y_{11},\, y_{5}y_{10}-y_{6}y_{11},\, 
y_{6}y_{8}-y_{5}y_{9},\, y_{3}y_{8}-y_{6}y_{9},\, \\
\nn
&y_{6}y_{7}-y_{8}y_{9},\, y_{5}y_{7}-y_{8}^2,\, y_{3}y_{7}-y_{9}^2,\, y_{3}y_{5}-y_{6}^2
\rangle \ ,
\end{align}
yielding
\begin{equation}
\kM := 
{\tiny
\left(\begin{array}{ccccccccc}
y_{3}& y_{5}& y_{6}& y_{7}& y_{8}& y_{9}& y_{10}& y_{11}& y_{12} \\ \hline
 0 & 0 & -1 & 0 & 0 & 1 & 0 & 1 & -1 \\
 0 & -1 & 0 & 0 & 1 & 0 & 0 & 1 & -1 \\
 0 & 0 & 0 & 1 & -1 & 0 & 0 & 1 & -1 \\
 -1 & 0 & 0 & 0 & 0 & 1 & 1 & 0 & -1 \\
 0 & 0 & -1 & 0 & 1 & 0 & 1 & 0 & -1 \\
 0 & 0 & 0 & 1 & 0 & -1 & 1 & 0 & -1 \\
 -1 & 0 & 1 & 0 & 0 & 0 & 1 & -1 & 0 \\
 0 & 1 & -1 & 0 & 0 & 0 & 1 & -1 & 0 \\
 0 & -1 & 1 & 0 & 1 & -1 & 0 & 0 & 0 \\
 1 & 0 & -1 & 0 & 1 & -1 & 0 & 0 & 0 \\
 0 & 0 & 1 & 1 & -1 & -1 & 0 & 0 & 0 \\
 0 & 1 & 0 & 1 & -2 & 0 & 0 & 0 & 0 \\
 1 & 0 & 0 & 1 & 0 & -2 & 0 & 0 & 0 \\
 1 & 1 & -2 & 0 & 0 & 0 & 0 & 0 & 0 \\
\end{array}\right) 
} \ .
\end{equation}
Subsequently, we find that
\begin{equation}
\kN = {\small \left(
\begin{array}{cccc}
 -2 & 0 & 0 & -1 \\
 0 & 0 & 0 & 1 \\
 -1 & 0 & 0 & 0 \\
 0 & -2 & 2 & 1 \\
 0 & -1 & 1 & 1 \\
 -1 & -1 & 1 & 0 \\
 -1 & 1 & 0 & 0 \\
 0 & 1 & 0 & 1 \\
 0 & 0 & 1 & 1 \\
\end{array}
\right) } \ .
\qquad
\begin{array}{c}
\includegraphics[width=7cm]{cy4Toric.jpg}
\end{array}
\end{equation}
First, the fact that we get four-vectors is reassuring since we have computed the geometry to be an affine Calabi--Yau fourfold.
We note that the vector $(-1,0,0,1)$ is perpendicular to the difference of every one of the vectors subtracted by the first one, $(-2,0,0,-1)$.
This means that the four-vectors are co-hyperplanar at height $1$, having their endpoint all lying in a hyperplane of dimension three.
This is the Calabi--Yau condition, as corroborated by the palindromic numerator of the Hilbert series.
Hence, we can plot the projection of the toric diagram into three dimensions, by, say, removing the first column of coordinates. This is included in the above.

Finally, we recognize the ideal in~\eqref{M_EW} to be also binomial, already in minimal presentation in~\eqref{EW3gen}.
Repeating the above, we find that
\begin{equation}
\kM =
{\scriptsize
\left(
\begin{array}{ccccccccc}
 0 & 0 & 0 & 0 & -1 & 1 & 0 & 1 & -1 \\
 0 & -1 & 1 & 0 & 0 & 0 & 0 & 1 & -1 \\
 0 & 0 & 0 & -1 & 0 & 1 & 1 & 0 & -1 \\
 0 & 0 & 0 & -1 & 1 & 0 & 1 & -1 & 0 \\
 -1 & 0 & 1 & 0 & 0 & 0 & 1 & 0 & -1 \\
 -1 & 1 & 0 & 0 & 0 & 0 & 1 & -1 & 0 \\
 0 & -1 & 1 & 0 & 1 & -1 & 0 & 0 & 0 \\
 -1 & 0 & 1 & 1 & 0 & -1 & 0 & 0 & 0 \\
 -1 & 1 & 0 & 1 & -1 & 0 & 0 & 0 & 0 \\
\end{array}
\right)}
\Rightarrow
\kN =
{\scriptsize
\left(
\begin{array}{ccccc}
 1 & 0 & 1 & 0 & 0 \\
 1 & 1 & 0 & 0 & 0 \\
 1 & 1 & 0 & 1 & 0 \\
 0 & 0 & 1 & 0 & 0 \\
 0 & 1 & 0 & 0 & 0 \\
 0 & 1 & 0 & 1 & 0 \\
 0 & -1 & 1 & -1 & 1 \\
 0 & 0 & 0 & -1 & 1 \\
 0 & 0 & 0 & 0 & 1 \\
\end{array}
\right)
} \ ;
\end{equation}
again, the rows of $\kN$ are five-vectors, as is needed for a five-fold.

\section{Affine Calabi--Yau Toric Varieties}\label{ap:toricCY}
In this appendix, we will clarify, in a rigorous manor, the meaning of ``toric Calabi--Yau'', and its relation to the property of being Gorenstein, and having palindromic numerator in the Hilbert series.
Importantly, we present the proof of the useful equivalent condition for a (possibly singular) toric variety to be Calabi--Yau: that the toric diagram be co-hyperplanar at height one. This extra height-one condition is often overlooked in the physics community --- where toric Calabi--Yau is often taken to mean coplanar toric diagram --- and needs to be emphasized. Indeed, whereas in the smooth case~\cite{Bouchard:2007ik}, this height one condition is redundant, it is crucial for the singular case, as are the ones discussed here and in the context of D-branes in AdS/CFT.

Throughout this section, we will make extensive use of the wonderful new text book on toric varieties~\cite{toricbook}.
When we say ``cone'', we mean a strongly convex rational polyhedral cone.
Let ${ N} = \IZ^d$, and ${ M} = \Hom_\IZ({ N}, \IZ) = \IZ^d$ its dual lattice.
To keep from getting confused, it is helpful to use this terminology, instead of simply writing $\IZ^d$ for the two dual lattices.

Let $\Sigma \subset { N}_\IR = \IR^d$ be a full-dimensional fan (full
dimensional means that the corresponding toric variety has ``no torus
factors'').
Let $$\phi = \{v_1 \ldots v_r\} : \IZ^r \longrightarrow { N}$$
be the matrix whose columns $v_i \in { N}$ are the $r$ rays of $\Sigma$.
The entries in each $v_i$ are integers, having greatest common divisor equal to one.

Let $X = X_\Sigma$ be the $d$-dimensional normal toric variety corresponding to $\Sigma$.
$X$ is determined by the following data: the $d \times r$ matrix of rays, $\phi$, and
the maximal cones of $\Sigma$.  Each cone can be thought of as a subset of $\{1, \ldots, r\}$,
i.e., we write $i \in \sigma$ to mean that the $i$th ray $v_i$ is an extremal ray of $\sigma$.

When the toric variety $X_\Sigma$ is smooth, then we have the following
theorem (\textit{cf}.~\cite{Bouchard:2007ik, reffert}).
\begin{theorem}\label{bouchard}
The {\em smooth} toric variety $X = X_\Sigma$ is Calabi--Yau if and only if
the endpoints of the extremal rays $v_1, \ldots, v_r$ all lie on an
affine hyperplane of the form:
  $$a_1 x_1 + \ldots + a_d x_d = 1,$$
where the $a_i$ are all {\em rational}.
\end{theorem}

Consider the following example.  $X = X_\sigma$ is the affine toric variety defined by  the cone $\sigma$, where $\sigma\check{}$ is the cone in ${ M}_\IR$ which is the convex hull of 
\begin{equation}\label{vero}
\{-1,-2,0\},\{1,0,0\},\{0,-1,0\},\{1,0,-2\},\{1,0,-1\},\{0,-1,-1\}.
\end{equation}
This is the cone in $\IC^5$ over the Veronese surface in $\IP^4$ discussed in Section \ref{toric}, in particular, the row of ${ N}$ in \eqref{Nvero}.

It is easy to check that the affine coordinate ring $A_\sigma$ is not Gorenstein, and therefore (see later in this appendix) is not Calabi--Yau either.
Using {\it Macaulay2}~\cite{mac} to compute this cone, we obtain that $\sigma$ has extremal rays 
$$\{0, -1, 0\}, \{0, 0, -1\}, \{2, -1, 1\}.$$
The endpoints of these rays do lie on an affine hyperplane: $\frac{x-y-z}{2}=1$.
This appears to be in contradiction with Theorem \ref{bouchard}.
We will show that it is not since $X_\sigma$ is not smooth: it is the cone over the Veronese surface and thus there is a singularity at the origin.

The main purpose of this appendix is to sketch a proof of the following extension of Theorem \ref{bouchard}, viz.,
\begin{theorem}~\label{thm:endpoints}
The toric variety $X_\Sigma$ is Calabi--Yau if and only if
the endpoints of the extremal rays $v_1, \ldots, v_r$ all lie on an
affine hyperplane of the form:
  $$a_1 x_1 + \ldots + a_d x_d = 1,$$
where the $a_i$ are all {\em integers}.
\end{theorem}

Notice that this theorem applies to the above example, and shows that it is not Calabi--Yau, since the coefficients of the affine hyperplane are rational, not integer.

In order to prove the statement, let us be careful with the definition of
Calabi--Yau.  As in the smooth case, we say that the toric variety $X =
X_\Sigma$ is {\em Calabi--Yau} if the canonical sheaf $\omega_X \cong
\OO_X$.  This implies that the Weil divisor $K_X = - D_1 - \ldots -
D_r$ is a Cartier divisor (i.e., locally, is generated by a single
equation), and in the class group $Cl(X)$, that $0 = - D_1 - D_2 - \ldots - D_r$.

We use the following well-known result, proved for example in Cox--Little--Schenck~\cite{toricbook}.
\begin{fact}[Theorem 4.1.3, page 172,~\cite{toricbook}]~\label{fact1}
The class group of $X$ is generated by $D_1, \ldots, D_r$, and
a presentation for this group is $\coker \phi^{T}$:
$$0 \longrightarrow { M} \longrightarrow \IZ^r \longrightarrow Cl(X) \longrightarrow 0.$$
\end{fact}

\bigskip
\noindent
{\bf Proof of Theorem~\ref{thm:endpoints}}

\medskip
\noindent
We wish to show that $X$ is Calabi--Yau exactly when there is a vector $m \in { M}$ such that $\langle m, v_i\rangle = 1$,
for all $i = 1, \ldots, r$ (this is the definition of the desired affine hyperplane).
Now,
$X$ is Calabi--Yau exactly when $0 \sim -D_1 - \ldots - D_r$, which is equivalent to this element being zero in the class
group, which is the same as saying that the vector $(-1,-1,\ldots,-1)$ is in the $\IZ$-span of the rows of $\phi$,
thanks to Fact~\ref{fact1}.  But to be in the $\IZ$-span of the rows is the same as the existence of an integer 
vector $m' \in \IZ^d = { M}$ such that $\langle m, v_i\rangle = -1$ for all $i$.  Taking $m = -m'$ gives our desired
hyperplane. \qed

\bigskip
Let us now restrict to the case of affine toric varieties.  We will show that the Calabi--Yau property in this case is just the Gorenstein-ness of the affine coordinate ring.  This is well-known, but we include the short proof for completeness.

\begin{theorem}~\label{thm:gor}
Let $X = X_\sigma$ be an affine normal toric variety. Then
$$\mbox{$X$ is Gorenstein} \iff \mbox{$X$ is Calabi--Yau}.$$
\end{theorem}

Recall that $X$ is called {\em Gorenstein} if the dualizing sheaf $\omega_X$ is a line bundle.  In terms of divisors, this means that $X_\sigma$ is Gorenstein exactly when $K_X = - D_1 - \ldots - D_r$ is a Cartier divisor.

We will need the following fact about when Weil divisors are Cartier on toric varieties $X_\Sigma$.

\begin{fact}[Theorem 4.2.8, page 181,~\cite{toricbook}]~\label{fact2}
A Weil divisor $D = \sum_{i=1}^r a_i D_i$ is Cartier if and only if for each 
maximal cone $\sigma \in \Sigma$, there is an integer vector $m_\sigma \in { M}$ such that
$\langle m_\sigma, v_i \rangle = -a_i$, for all $i \in \sigma$ (recall we are thinking of
$\sigma$ as a subset of the indices $\{1, \ldots, r\}$).
\end{fact}

\bigskip
\noindent
{\bf Proof of Theorem~\ref{thm:gor}}
\medskip

\noindent
If $X$ is Calabi--Yau, then as noted above, $K_X$ is Cartier, hence $X$ is Gorenstein.  Conversely, if $X$ is Gorenstein, then $- D_1 - \ldots - D_r$ is Cartier.  The just quoted fact then implies that
there is a vector $m_\sigma \in { M}$ (there is only one cone in this case), such that $\langle m, v_i\rangle = 1$
for all $i = 1..r$.  But this is just the equivalent condition proved above, showing that $X_\sigma$ is Calabi--Yau. 
\qed

It is also well known that $X_\sigma$ is Gorenstein exactly when the
affine coordinate ring $A_\sigma$ is Gorenstein.  One way to see this
is the following.  Recall that the 
canonical module of $X_\sigma$ and of the coordinate ring $A_\sigma$ are the same,
and that $\omega_X \subset A_\sigma$ is the subset generated by the
monomials in the interior of $\check{\sigma}$.
$A_\sigma$ is Gorenstein exactly when this ideal is generated by a single element.
This is
equivalent to the canonical divisor being the divisor of an invariant
rational function on $X_\sigma$ and therefore the canonical divisor is
principal.  But any principal divisor is Cartier, so the canonical divisor is Cartier, and therefore
$X_\sigma$ is Gorenstein. 
The
other direction is simpler: Given that $\omega_X$ is Cartier, this
means that there exists an $m \in { M}$ such that $\langle m, v_i \rangle
= 1$, for all $v_i$.  The canonical module is
generated by the corresponding monomial, therefore $A_\sigma$ is
Gorenstein.

\subsection{Illustrative Examples}
To illustrate the foregoing discussions, let us take two concrete examples:
the Veronese from the previous 
appendix, and the quotient $\IC^3/\IZ_2\times\IZ_2$.  
The first is not Calabi--Yau, but the second example is.
To be completely explicit, we will use {\it Macaulay2} \cite {mac} to analyze them, and include all the relevant code for reference. 

To simplify the examples below, we first load the following {\it Macaulay2} code.  This  is in the file {\tt toric-calabi-yau.m2} included below.

{\scriptsize
\begin{verbatim}
-- file: toric-calabi-yau.m2

-- load the packages we will use:
needsPackage "FourTiTwo"
needsPackage "Polyhedra"
needsPackage "NormalToricVarieties"

-- Compute and display a hyperplane passing through the endpoints of the given rays
-- if such a hyperplane exists.
findHyperplane = method()
findHyperplane Cone := (C) -> findHyperplane entries transpose rays C
findHyperplane List := (rays) -> (
    M := (matrix rays) | (matrix {#rays:{1}});
    Z := syz M;
    if numColumns Z == 0 then return "no hyperplane exists";
    if numColumns Z > 1 then error "original cone is not full dimensional";
    Z = flatten entries Z;
    if Z#-1 < 0 then Z = -Z;
    -- construct the hyperplane from Z
    G := sum apply(#Z - 1, i -> (-Z_i) * expression(x_(i+1)));
    G/Z#-1 == 1
    )

-- Compute the ideal of the affine toric variety.  
-- This particular function uses the 4ti2 package, which is overkill for the
-- examples in this section, but is useful on large examples.
toricIdeal = method()
toricIdeal(Cone, Symbol) := (C, x) -> (
    C' := dualCone C;
    H := matrix {hilbertBasis C'};
    A := transpose matrix H;
    ncols := numColumns H;
    R = QQ[x_1..x_ncols];
    trim toricMarkov(H,R)
    )
\end{verbatim}
}

\paragraph{Example 1: Veronese: }
The first example is from the previous appendix.
We start with the vectors from Equation~\eqref{vero}.
Let $C \subset { N}_\IR$ be the dual cone to the cone $C'$
spanned by the columns of the matrix $m$:

{\scriptsize
\begin{verbatim}
i1 : load "toric-calabi-yau.m2"

i2 : m = transpose matrix {{-1,-2,0},{1,0,0},{0,-1,0},{1,0,-2},{1,0,-1},{0,-1,-1}}

o2 = | -1 1 0  1  1  0  |
     | -2 0 -1 0  0  -1 |
     | 0  0 0  -2 -1 -1 |

              3        6
o2 : Matrix ZZ  <--- ZZ

i3 : C' = posHull m

o3 = {ambient dimension => 3           }
      dimension of lineality space => 0
      dimension of the cone => 3
      number of facets => 3
      number of rays => 3

o3 : Cone

i4 : C = dualCone C'

o4 = {ambient dimension => 3           }
      dimension of lineality space => 0
      dimension of the cone => 3
      number of facets => 3
      number of rays => 3

o4 : Cone
\end{verbatim}
}
\noindent
The extremal rays of the cone $C \subset { N}_\IR$ are the columns of the following matrix.

{\scriptsize
\begin{verbatim}
i5 : rays C

o5 = | 0  0  2  |
     | -1 0  -1 |
     | 0  -1 1  |

              3        3
o5 : Matrix ZZ  <--- ZZ
\end{verbatim}
}
\noindent
The hyperplane is ``at height 2'', so the affine toric variety $X$
corresponding to the cone $C$ is not Calabi--Yau:

\newpage

{\scriptsize
\begin{verbatim}
i6 : findHyperplane C

     x  - 2x  - 2x
      1     2     3
o6 = -------------- == 1
            2

o6 : Expression of class Equation
\end{verbatim}
}
\noindent
The ideal of this toric variety is in a polynomial ring which has one variable
for each Hilbert basis generator of the dual cone of $C$.  The corresponding ideal
is the Veronese, and its Hilbert series is not palindromic, as expected, since
this variety is not Calabi--Yau.

{\scriptsize
\begin{verbatim}
i7 : hilbertBasis dualCone C

o7 = {| 0  |, | 1  |, | 1  |, | -1 |, | 0  |, | 1 |}
      | -1 |  | 0  |  | 0  |  | -2 |  | -1 |  | 0 |
      | -1 |  | -1 |  | -2 |  | 0  |  | 0  |  | 0 |

o7 : List

i8 : I = toricIdeal(C,symbol x)

             2                                    2                       2
o8 = ideal (x  - x x , x x  - x x , x x  - x x , x  - x x , x x  - x x , x  - x x )
             5    4 6   2 5    1 6   2 4    1 5   2    3 6   1 2    3 5   1    3 4

o8 : Ideal of R

i9 : reduceHilbert hilbertSeries I

      1 + 3T
o9 = --------
            3
     (1 - T)

\end{verbatim}
}
\noindent
We now use the {\tt NormalToricVarieties} {\it Macaulay2} package, written by Greg Smith
and included with {\it Macaulay2},
to analyze this variety in a somewhat higher level fashion.  First, the function
{\tt normalToricVariety} expects a list of rays of all of the cones in the fan of a toric variety,
as well as a list of list of indices, indicating which rays correspond to maximal cones
in the fan.

{\scriptsize
\begin{verbatim}
i10 : raysC = entries transpose rays C

o10 = {{0, -1, 0}, {0, 0, -1}, {2, -1, 1}}

o10 : List

i11 : X = normalToricVariety(raysC, {{0,1,2}})

o11 = X

o11 : NormalToricVariety
\end{verbatim}
}
\noindent
By the results of this appendix, we see that the affine toric variety $X$ is not Calabi--Yau.

{\scriptsize
\begin{verbatim}
i12 : isSmooth X

o12 = false

i13 : KX = toricDivisor X -- the (toric) canonical divisor on X

o13 = - D  - D  - D
         0    1    2

o13 : ToricDivisor on X

i14 : isCartier KX -- not Cartier, therefore X is not Calabi-Yau

o14 = false
\end{verbatim}
}
\noindent
Finally, let us desingularize $X$, which corresponds to subdividing 
the cone to smooth simplicial cones.  In this case, one ray is added, resulting in
three maximal cones.  The resulting smooth toric variety $Y$ is not Calabi--Yau.

{\scriptsize
\begin{verbatim}
i15 : Y = makeSmooth X

o15 = Y

o15 : NormalToricVariety

i16 : raysY = rays Y

o16 = {{0, -1, 0}, {0, 0, -1}, {2, -1, 1}, {1, -1, 0}}

o16 : List

i17 : conesY = max Y

o17 = {{0, 1, 3}, {0, 2, 3}, {1, 2, 3}}

o17 : List

i18 : findHyperplane raysY -- But Y is not Calabi-Yau

o18 = no hyperplane exists

i19 : isCartier toricDivisor Y -- However, Y is smooth, therefore Gorenstein

o19 = true
\end{verbatim}
}

\medskip
\paragraph{Example 2: $\IC^3/\IZ_k\times\IZ_k$: }
We analyze the quotient $X = \IC^3/\IZ_k\times\IZ_k$, for $k=2$,
although we could set $k$ to other values as well.
As we see during the example, $X$ is Calabi--Yau, as expected.

{\scriptsize
\begin{verbatim}
i20 : k = 2

i21 : m = transpose matrix {{k,-1,-1},{0,0,1},{0,1,0},{1,0,0}}

o21 = | 2  0 0 1 |
      | -1 0 1 0 |
      | -1 1 0 0 |

               3        4
o21 : Matrix ZZ  <--- ZZ

i22 : C = dualCone posHull m

o22 = {ambient dimension => 3           }
       dimension of lineality space => 0
       dimension of the cone => 3
       number of facets => 3
       number of rays => 3

o22 : Cone
\end{verbatim}
}
\noindent
The cone $C$ is the cone whose extremal rays are the columns of the following matrix.

{\scriptsize
\begin{verbatim}
i23 : rays C

o23 = | 1 1 1 |
      | 0 2 0 |
      | 0 0 2 |

               3        3
o23 : Matrix ZZ  <--- ZZ
\end{verbatim}
}
\noindent
In this case, it is clear that the endpoints of these rays lie on the plane $x_1 = 1$, and so
the corresponding toric variety is Calabi--Yau.

{\scriptsize
\begin{verbatim}
i24 : findHyperplane C -- the toric variety corresponding to this cone is Calabi-Yau

o24 = x  == 1
       1

o24 : Expression of class Equation
\end{verbatim}
}
\noindent
The ideal defining the toric variety is a toric hypersurface.
As such affine toric hypersurfaces are all Calabi--Yau, this gives an independent
confirmation that $X$ is Calabi--Yau.

{\scriptsize
\begin{verbatim}
i25 : I = trim toricIdeal(C,symbol x) -- singular toric hypersurface, so Calabi-Yau

                      2
o25 = ideal(x x x  - x )
             2 3 4    1

o25 : Ideal of R

i26 : raysC = entries transpose rays C

o26 = {{1, 0, 0}, {1, 2, 0}, {1, 0, 2}}

o26 : List

i27 : X = normalToricVariety(raysC, {{0,1,2}})

o27 = X

o27 : NormalToricVariety

i28 : isSmooth X

o28 = false

i29 : KX = toricDivisor X -- the (toric) canonical divisor on X

o29 = - D  - D  - D
         0    1    2

o29 : ToricDivisor on X

i30 : isCartier KX -- Cartier, therefore X is Calabi-Yau

o30 = true
\end{verbatim}
}
\noindent
The desingularization $Y$ of $X$ has 6 rays, and the original cone has been subdivided
into 4 smaller simplicial cones.

{\scriptsize
\begin{verbatim}
i31 : Y = makeSmooth X

o31 = Y

o31 : NormalToricVariety

i32 : raysY = rays Y

o32 = {{1, 0, 0}, {1, 2, 0}, {1, 0, 2}, {1, 1, 0}, {1, 0, 1}, {1, 1, 1}}

o32 : List

i33 : conesY = max Y

o33 = {{0, 3, 4}, {1, 3, 5}, {2, 3, 4}, {2, 3, 5}}

o33 : List
\end{verbatim}
}
\noindent
The desingularization $Y$ is Calabi--Yau, and $Y$ is Gorenstein, since it is nonsingular.
{\scriptsize
\begin{verbatim}
i34 : findHyperplane raysY -- Y is Calabi-Yau too

o34 = x  == 1
       1

o34 : Expression of class Equation

i35 : isCartier toricDivisor Y -- Y is smooth, therefore Gorenstein

o35 = true
\end{verbatim}
}

If we were to start with the {\it dual} of $C$, instead of $C$, 
we would obtain an affine toric variety which is not Calabi--Yau,
which would in fact be the cone over the $k$-uple embedding of $\IP^2$.

\end{document}